\documentclass[twoside,11pt]{article}

\usepackage{blindtext}

\usepackage[abbrvbib, preprint]{dmlr2e}

\usepackage{graphicx}
\usepackage[utf8]{inputenc} 
\usepackage[T1]{fontenc}    
\usepackage{hyperref}       
\usepackage{url}            
\usepackage{booktabs}       
\usepackage{amsfonts}       
\usepackage{nicefrac}       
\usepackage{microtype}      
\usepackage{xcolor}         
\usepackage{amsmath,amssymb,amsfonts}
\usepackage{caption,subcaption}
\usepackage{pifont}
\usepackage{multirow}
\usepackage{tikz}

\newcommand{\cmark}{\ding{51}}%
\newcommand{\xmark}{\ding{55}}%

\usepackage{lastpage}
\dmlrheading{23}{2026}{1-\pageref{LastPage}}{12/25; Revised 09/26}{10/26}{21-0000}{Tiantian Feng, Jihwan Lee, Anfeng Xu, Yoonjeong Lee, Thanathai Lertpetchpun, Xuan Shi, Helin Wang, Thomas Thebaud, Laureano Moro-Velazquez, Dani Byrd, Najim Dehak, Shrikanth Narayanan} 
\def\openreview{\url{https://openreview.net/forum?id=qmGQoZfkEN}} 

\ShortHeadings{Vox-Profile}{Tiantian Feng et al.}
\firstpageno{1}

\begin{document}

\title{Vox-Profile: A Speech Foundation Model Benchmark for Characterizing Diverse Speaker Traits and Speech Properties}

\author{
  \name \hspace{-1.5mm}Tiantian Feng$^1$ \email tiantiaf@usc.edu \\
  \name Jihwan Lee$^1$ \email jihwan@usc.edu  \\
  \name Anfeng Xu$^1$ \email anfengxu@usc.edu \\
  \name Yoonjeong Lee$^1$  \email yoonjeol@usc.edu \\
  \name Thanathai Lertpetchpun$^1$ \email lertpetc@usc.edu \\
  \name Xuan Shi$^1$ \email xuanshi@usc.edu \\
  \name Helin Wang$^2$ \email hwang258@jhu.edu  \\
  \name Thomas Thebaud$^2$ \email tthebau1@jhu.edu \\
  \name Laureano Moro-Velazquez$^2$ \email laureano@jhu.edu \\
  \name Dani Byrd$^1$ \email dbyrd@usc.edu \\
  \name Najim Dehak$^2$ \email ndehak3@jhu.edu \\
  \name Shrikanth Narayanan$^1$ \email shrikann@usc.edu \\
  \\
  \addr $^1$University of Southern California \\
  \addr $^2$Johns Hopkins University \\
}

\editor{Peter Mattson}

\maketitle

\begin{abstract}
We introduce \texttt{Vox-Profile}, a comprehensive benchmark to characterize rich speaker traits and speech properties using speech foundation models. Unlike prior work that targets a single dimension of speaker traits, \texttt{Vox-Profile} provides holistic and multi-dimensional profiles spanning both static speaker traits (e.g., age, sex, accent, voice quality) and dynamic speech traits (e.g., emotion, speech flow). This benchmark is grounded in speech science and linguistics, developed with domain experts to accurately index speaker and speech characteristics. We report benchmark experiments using over 15 publicly available speech datasets and several widely used speech foundation models that target various static and dynamic speaker and speech properties.
In addition to benchmark experiments, we showcase several downstream applications supported by \texttt{Vox-Profile}. First, we show that \texttt{Vox-Profile} can augment existing speech recognition datasets to analyze ASR performance variability.
\texttt{Vox-Profile} is also used as a tool to evaluate the performance of speech generation systems. Finally, we assess the quality of our automated profiles through comparison with human evaluation and show convergent validity. \texttt{Vox-Profile} is publicly available at: \href{https://github.com/tiantiaf0627/vox-profile-release}{https://github.com/tiantiaf0627/vox-profile-release}.

\end{abstract}


\section{Introduction}

Even brief segments of voice carry a wealth of information about a speaker, including their age, sex, accent, emotion, and their physical environment and social context. The ability to accurately predict these attributes from speech has significant value in advancing a wide range of speech technologies. For example, conversational digital assistants like Alexa or Siri can adapt responses based on a user's perceived emotional state. Automatic speech recognition systems (ASR) \citep{prabhavalkar2023end} can improve transcription accuracy by modeling and adapting to different accents. Despite these promising applications, much of the progress in speech modeling has been centered on conventional tasks such as ASR for transcribing spoken language, speaker diarization \citep{park2022review} for tracking who is speaking when, and speech enhancement \citep{michelsanti2021overview} for improving audio speech quality. In contrast, there is limited work in systematically modeling and predicting diverse speaker and speech traits.

Much of the existing research characterizing speakers has focused on learning speaker embeddings, such as $x$-vectors~\citep{snyder2018x}, which capture the unique vocal traits of individual speakers. These embeddings emerged as foundational components in many applications, most notably in speaker recognition and verification systems. While significant progress has been made in using speaker embeddings to capture identity-specific vocal features, relatively little work has focused on comprehensive modeling of other human-interpretable speaker traits, such as accent and voice quality (or timbre). Recent work has shown that robust inclusion of such traits holds great promise for downstream applications, like style-prompted speech generation models. For example, the \texttt{ParaSpeechCaps}~\citep{diwan2025scaling} dataset provides large-scale human annotations of speaker traits to facilitate the development of high-performing text-to-speech (TTS) models capable of generating speech with different speaking styles.

Although several studies have explored the modeling of speaker traits such as accents \citep{zuluaga2023commonaccent} or age \citep{burkhardt2023speech}, each study designs its unique labeling choice in the classification, leading to inconsistencies across the literature. As a result, a unified framework for defining, categorizing, and modeling different speaker or speech traits is lacking. In this work, we propose \texttt{Vox-Profile}, one of the first benchmarking efforts that systematically evaluates rich multi-dimensional speaker and speech traits from English-speaking voices. Specifically, \texttt{Vox-Profile} experiments with over 15 publicly available datasets to predict static traits (e.g., age, sex, accent, and voice quality) and dynamic traits (e.g., speech emotion, speech flow) in different recording conditions and elicitation settings (e.g., read, spontaneous, and conversational speech). Our benchmark covers popular speech models, including \texttt{HuBERT}~\citep{hsu2021hubert}, \texttt{WavLM}~\citep{chen2022wavlm}, \texttt{ECAPA-TDNN}~\citep{desplanques2020ecapa}, and \texttt{Whisper} family~\citep{radford2023robust}.

Our proposed \texttt{Vox-Profile} benchmark exhibits multifold merits compared to prior works: (1) Unlike previous work on modeling limited speaker traits or speech properties, \texttt{Vox-Profile} provides a holistic characterization of an extensive list of traits from speech, including age, sex, accents, emotions, voice quality, speech flow, and expressiveness. (2) Instead of solely focusing on computational modeling, \texttt{Vox-Profile} integrates linguistically informed labeling choices to better address the inconsistent label choices of speaker traits across datasets. (3) The broad utility of \texttt{Vox-Profile} is established through investigating several potential applications, including analyzing speech model performance, evaluating speech generation systems, and automatically tagging speaking styles. Both machine and human evaluations indicate that \texttt{Vox-Profile} offers reliable estimations of speaker and speech traits.

\begin{table}
    \centering

    \caption{Comparison of Vox-Profile with existing literature in modeling speaker and speech traits.}
    \vspace{0.5mm}
    \resizebox{\linewidth}{!}{
    \begin{tabular}{lcccccccc}

        \toprule
         & 
        \multirow{2}{*}{\textbf{Age}} & 
        \multirow{2}{*}{\textbf{Sex}} & 
        \multirow{2}{*}{\textbf{Accent}} & 
        \multicolumn{1}{c}{\textbf{Categorized}} & 
        \multicolumn{1}{c}{\textbf{Arousal}} & 
        \multicolumn{1}{c}{\textbf{Voice}} &
        \multicolumn{1}{c}{\textbf{Speech}} &
        \multicolumn{1}{c}{\textbf{Speech}}
        \\

        & & & & 
        \multicolumn{1}{c}{\textbf{Emotion}} &
        \multicolumn{1}{c}{\textbf{Valence}} & 
        \multicolumn{1}{c}{\textbf{Quality}} &
        \multicolumn{1}{c}{\textbf{Flow}} &
        \multicolumn{1}{c}{\textbf{Expressiveness}}
        \\
        \midrule

        VoxCeleb Enrichment \citep{hechmi2021voxceleb} & 
        \cmark & 
        \cmark & 
        \xmark & 
        \xmark & 
        \xmark &
        \xmark &
        \xmark &
        \xmark
        \\
        
        CommonAccent \citep{zuluaga2023commonaccent} & 
        \xmark & 
        \xmark & 
        \cmark & 
        \xmark & 
        \xmark &
        \xmark &
        \xmark &
        \xmark
        \\

        GLOBE \citep{wang2024globe} & 
        \cmark & 
        \cmark & 
        \cmark & 
        \xmark & 
        \xmark &
        \xmark &
        \xmark &
        \xmark
        \\

        ParaSpeechCaps \citep{diwan2025scaling} & 
        \xmark & 
        \cmark & 
        \cmark & 
        \cmark & 
        \xmark &
        \cmark &
        \xmark &
        \cmark
        \\

        \midrule
        Vox-Profile (Ours) & 
        \cmark & 
        \cmark & 
        \cmark & 
        \cmark & 
        \cmark &
        \cmark &
        \cmark &
        \cmark
        \\

        \bottomrule

    \end{tabular}
    }
    \label{tab:related_work}
\end{table}

\section{Related Work}

\paragraph{Static Trait Modeling.} We summarize several related representative works in Table~\ref{tab:related_work}. Age and sex prediction are long-standing tasks in speaker trait modeling. Burkhardt et al.\citep{burkhardt2023speech} introduced a benchmark framework that uses several publicly available datasets for age and sex prediction based on the transformer. In a related effort, Yang et al.\citep{yang2025demographic} investigated the effectiveness of \texttt{WavLM} embeddings for age and sex prediction, showing that self-supervised speech representation can achieve higher performances compared to i-vector features \citep{dehak2010front}. Apart from modeling age and sex attributes from speech, there has been growing interest in predicting speaker accents. For example, \texttt{CommonAccent}~\citep{zuluaga2023commonaccent} introduced a benchmark for accent classification using samples from Mozilla Common Voice~\citep{ardila2020common}. It reported strong performance in recognizing regional varieties of English, such as American, Canadian and British English, using the \texttt{ECAPA-TDNN} speaker embedding model. GLOBE \citep{wang2024globe} is a similar effort that develops classifiers using \texttt{HuBERT} pre-trained models on Common Voice to predict accent as well as sex and age.

\paragraph{Dynamic Trait Modeling.} Several works have also focused on modeling and characterizing dynamic speaker traits, such as speech emotion and fluency. One of the early efforts in supporting speech emotion recognition (SER) is the release of the IEMOCAP dataset \citep{busso2008iemocap}, which provides annotated recordings of acted dialogues with fine-grained emotional labels to develop emotion recognition systems. Several datasets have since been designed to support more naturalistic emotion modeling, including the MSP-Podcast~\citep{lotfian2017building}, MELD~\citep{poria2019meld}, CREMA-D~\citep{cao2014crema}, and CMU-MOSEI~\citep{zadeh2018multimodal} datasets. 
Recent advances in speech foundation models have also led to the development of high-performing SER systems. Leveraging pre-trained representations from models such as \texttt{WavLM} and \texttt{Whisper}, researchers have demonstrated improvements over traditional SER approaches~\citep{wagner2023dawn, feng2023peft}. Likewise, a growing body of research focused on predicting speech fluency to develop robust language assessment systems. For example, SEP-28K~\citep{lea2021sep} presents a notable effort to release large-scale speech data annotated for disfluencies, including conditions such as sound blocks.

\section{Labeling Choice of Speaker Traits and Speech Properties}

Unlike prior benchmarks focused on general speech modeling tasks~\citep{yang2021superb}, \texttt{Vox-Profile} offers a structured benchmarking framework for systematically predicting a broad range of speaker traits and speech properties. In developing this automated evaluation pipeline, we observed considerable inconsistency across existing studies in how speaker traits and speech properties are categorized, defined, and modeled. One of the closest prior efforts is \texttt{ParaSpeechCaps}~\citep{diwan2025scaling}, which proposes a labeling choice that models ``intrinsic'' (e.g., accent) and ``situational'' (e.g., emotion) traits. However, its categorization remains underspecified and difficult to reconcile with conventions based on empirical foundations in speech science literature. 
For example, while \texttt{ParaSpeechCaps} includes a broad set of emotional descriptors, these labels deviate from standard categorical and dimensional emotion frameworks commonly used in speech and language research, creating challenges when aligning datasets or comparing across studies. 

\begin{table}
    
    \centering

    \caption{The labeling scheme of speaker traits and speech properties in \texttt{Vox-Profile} benchmark. The green, blue, and violet indicate speaker accents from North America, the British Isles, and other regions or language backgrounds, respectively. For voice quality, the colors green, blue, violet, olive, and orange represent dimensions of speaker pitch, rhythm, clarity, voice texture, and volume defined by \texttt{ParaSpeechCaps}. Finally, the blue color denotes labels related to speech disfluencies in the speech flow category.}

    \resizebox{\linewidth}{!}{
    \begin{tabular}{lcccccccc}

        \toprule
        \multicolumn{1}{c}{\textbf{Category}} & 
        \multicolumn{1}{c}{\textbf{Traits}} & 
        \multicolumn{1}{c}{\textbf{Labeling Scheme}} & 
        \multicolumn{1}{c}{\textbf{Datasets}} 
        \\
        \midrule

        \multirow{14}{*}{\textbf{\shortstack{Static\\Traits}}} & 
        \multirow{2}{*}{\textbf{Speaker Sex}} & 
        \multirow{2}{*}{Male; Female} & TIMIT\citep{garofolo1993darpa}; VoxCeleb\citep{nagrani2017voxceleb}; 
        \\
        & & & CommonVoice\citep{ardila2020common}
        \\
        \cmidrule(lr){2-4}
        
        & \multirow{2}{*}{\textbf{Speaker Age}} & 
        Young adults (<30 Years); Adults (30-60 years); & {TIMIT\citep{garofolo1993darpa}; VoxCeleb\citep{nagrani2017voxceleb};}
        \\

         & & 
        Senior adults (>60 years) & CommonVoice\citep{ardila2020common}
        \\
        \cmidrule(lr){2-4}

         & \multirow{6}{*}{\textbf{Speaker Accent}} & 
        \textcolor{teal}{North America}, \textcolor{blue}{English} & CommonVoice\citep{ardila2020common};
        \\

        & & 
        \textcolor{teal}{North America}, \textcolor{blue}{Welsh, Scottish} & EdAcc \citep{sanabria2023edinburgh}; CSLU-FAE
        \\

         & & 
        \textcolor{blue}{Northern Irish, Irish,} \textcolor{violet}{Germanic} &  
        British Isles \citep{demirsahin2020open}; TIMIT~\citep{garofolo1993darpa} \\

         & & 
        \textcolor{violet}{Romance, Slavic, Semitic} &  
         L2-ARCTIC~\citep{zhao2018l2}; VoxPopuli~\citep{wang2021voxpopuli}; \\

         & & 
        \textcolor{violet}{Oceania, South Africa, East Asia} &
        Fair-Speech \citep{veliche2024towards}; ESLTTS \citep{wang2024usat} \\

         & & 
        \textcolor{violet}{Southeast Asia, South Asia, Other} &  
        Hispanic-English~\citep{hispanic_eng}; Nigerian-English~\citep{nigerian_eng}\\

        \cmidrule(lr){2-4}
        
        & \multirow{5}{*}{\textbf{Voice Quality}} & 
        \textcolor{teal}{Shrill, Nasal, Deep;} \textcolor{blue}{Singsong, Pitchy, Flowing,} & \multirow{5}{*}{ParaSpeechCaps \citep{diwan2025scaling}}
        \\

        & & 
        \textcolor{blue}{Monotone, Staccato, Punctuated, Enunciated,} & \\

        & & 
        \textcolor{blue}{Hesitant;} \textcolor{violet}{Crisp, Slurred, Lisp, Stammering;} & \\

        & & 
        \textcolor{olive}{Silky, Husky, Raspy, Guttural, Vocal-fry;} &
        \\

        & & 
        \textcolor{orange}{Booming, Authoritative, Loud, Hushed, Soft} & 
        \\

        \midrule
        
        \multirow{8}{*}{\textbf{\shortstack{Dynamic\\Traits}}} & 
        \textbf{Categorical} & 
        Neutral, Happy, Sad, Angry, Contempt & \multirow{2}{*}{MSP-Podcast \citep{lotfian2017building}}
        \\

        & \textbf{Emotion} & 
        Fear, Disgust, Surprise, Other & 
        \\
        \cmidrule(lr){2-4}

        & \textbf{Arousal/Valence} & 
        \textcolor{teal}{0-1(Calm->Active)} / \textcolor{blue}{0-1(Negative->Positive)} & MSP-Podcast \citep{lotfian2017building}
        \\
        \cmidrule(lr){2-4}
        
        & 
        \multirow{2}{*}{\textbf{Speech Flow}} & 
        \textcolor{teal}{Fluent, Disfluent} \textcolor{blue}{(Prolongation, Word Repetition} & {Sep-28K \citep{lea2021sep}}
        \\
        & & 
        \textcolor{blue}{Sound Repetition, Block, Interjection)} & FluencyBank \citep{ratner2018fluency}
        \\
        \cmidrule(lr){2-4}
        
        & \textbf{Speech} & Animated, Laughing, Passive & \multirow{2}{*}{ParaSpeechCaps \citep{diwan2025scaling}}
        \\
        
        & \textbf{Expressiveness} & Whispered, Enunciated & 
        \\

        \bottomrule

    \end{tabular}
    }
    \label{tab:taxonomy}
\end{table}

Many existing approaches in speech trait modeling lack grounding in speech science and linguistics, which often leads to ambiguous or ill-posed problem formulations. A prominent example is speaker age modeling, which is frequently treated as a regression task targeting a speaker’s exact chronological age~\citep{si2022towards, Dareeniassp25}. However, this framing overlooks numerous confounding variables such as biological sex, vocal anatomy, and speaking habits that affect vocal aging. Crucially, it misaligns with how age is typically perceived (and experienced) from speech: human listeners estimate age in broad {\em intervals} rather than in precise years~\citep{schotz2006perception, skoog2015can}. Similarly, accent classification is often based on a speaker's nationality, as in \texttt{ParaSpeechCaps}, even though nationality is not reliably inferable from speech alone, particularly in multilingual or diasporic contexts~\citep{rubin1992nonlanguage,humayun2024review}. To address these challenges, we introduce a linguistically principled labeling choice in \texttt{Vox-Profile}, designed to be generalizable across diverse speech datasets and conditions.

We define two broad categories of speech traits to structure the benchmark: (1) Static speaker traits capture relatively stable characteristics of the speaker: age, sex, accent, and voice quality; and (2) dynamic speech traits reflect context-dependent aspects of speech that can vary across situations, including emotion, speech expressiveness, and speech flow. The full labeling choices are summarized in Table~\ref{tab:taxonomy}, with the description of all datasets used in this benchmark and data splits for the benchmark experiment detailed in Appendix~\ref{appendix-datasets}.

\subsection{Static Speaker Traits} 

\paragraph{Speaker Sex} Biological sex as reflected in voice is relatively well-defined in the literature. We follow standard practice by modeling it as a binary classification (male vs. female), while recognizing that, in practice, acoustic parameters of voice quality do not always map cleanly onto binary categories, particularly in the case of young children or speakers with atypical vocal profiles. 

\paragraph{Speaker Age} As discussed earlier, many existing studies treat age estimation as a regression task, aiming to predict a speaker's exact chronological age from speech~\citep{si2022towards, Dareeniassp25}. We argue that this formulation is ill-posed: perceived vocal age is shaped by multiple interacting factors and is more naturally categorized in intervals than estimated precisely. Instead, we propose classifying speakers into three broad age groups: young adults (18-30 years), adults (30-60 years), and senior adults (>60 years). Speech science research~\citep{reubold2010vocal} indicates that the vocal characteristics typically stabilize in early adulthood (approximately 18 years), with relatively consistent voice characteristics across the adult decades. In contrast, age-related vocal changes, linked to hormonal, physiological, and health-related factors, become increasingly salient from around age 60 years onward.

\paragraph{Speaker Accent} An accent refers to a distinctive way of pronouncing a language, typically shaped by a speaker's regional origin and linguistic background. However, the classification of accents has varied widely across the literature with inconsistent labeling schemes and taxonomic assumptions~\citep{wang2024globe,zuluaga2023commonaccent, zhong2025accentbox}. In this work, we introduce a unified and scalable labeling choice for accent classification, designed to integrate multiple mainstream English-accent datasets under a coherent framework.  Focusing on English-speaking populations, we first organize accents into three broad regional groups: North American, British Isles, and other regions or other language backgrounds (e.g., Oceania, South Asia, and Africa). Within the British Isles, we further distinguish specific regional varieties: English (England), Scottish, Northern Irish, Welsh, and Irish. To better capture systematic cross-linguistic influences on English pronunciation, we also group certain English accents by the language family of the speakers' first language (L1). This includes Germanic (e.g., German, Dutch), Slavic (e.g., Russian, Polish), Romance (e.g., Spanish, French, Italian), and Semitic (e.g., Arabic, Hebrew) language backgrounds. This typological grouping reflects the fact that related L1s often give rise to similar segmental and prosodic transfer patterns in second-language (L2) English, which do not always align clearly with national or geographic borders. Asian accents are organized by region, East Asia, South Asia, and Southeast Asia, reflecting common typological and areal features. Additional categories include Oceania and South Africa, representing other major English-speaking regions. Due to underrepresentation in existing datasets, some language backgrounds are not able to be classified into these categories reliably. For example, accents associated with Uralic languages as L1s (e.g., Finnish) or with regions of Africa outside South Africa are grouped under the general ``Other'' label. This labeling choice enables the integration of $11$ publicly available datasets, each with distinct accent labeling conventions, into a unified benchmark for training and evaluating accent classification models.

\paragraph{Voice Quality} Only a few publicly available datasets offer detailed annotations suitable for modeling voice quality. For this benchmark, we adopt \texttt{ParaSpeechCaps}~\citep{diwan2025scaling}, the dataset with the most extensive human-annotated labels currently available in this category. It defines voice quality across five perceptual dimensions: pitch, voice texture, volume, clarity, and rhythm, each reflecting a distinct aspect of how a speaker's voice is perceived beyond segmental content. The specific labels associated with each dimension are provided in Table~\ref{tab:taxonomy}.

\subsection{Dynamic Speech Traits}

\paragraph{Categorical Emotion} There is an extensive list of SER datasets, including IEMOCAP~\citep{busso2008iemocap}, MELD~\citep{poria2019meld}, CMU-MOSEI~\citep{zadeh2018multimodal}, and CREMA-D~\citep{cao2014crema}, each employing distinct emotion labeling schemes. Among them, we find the MSP-Podcast dataset~\citep{lotfian2017building} provides the largest number of speech emotion samples recorded in naturalistic settings. Therefore, we adopt its emotion labeling scheme, including categories neutral, happy, sad, angry, fear, disgust, contempt, surprise, non-agreement, and others.

\paragraph{Arousal/Valence} Arousal and valence are also widely used to describe emotional states in speech~\citep{kuppens2013relation}. We define both arousal and valence as continuous variables scaled from 0 to 1. \textit{Arousal} represents the intensity of the emotion, with higher scores indicating greater activation or energy.  \textit{Valence} reflects emotional polarity, with lower scores corresponding to more negative affective states (e.g., sadness) and higher values to more positive (e.g., happiness).

\paragraph{Speech Flow} Speech flow has gained growing attention in recent years as a dimension of fluency in spoken language~\citep{ratner2018fluency}. One of the most notable resources is the Sep-28K dataset~\citep{lea2021sep}, which annotates disfluencies at scale. Following this foundation, we categorize speech flow into two: fluent and disfluent. Disfluency is further defined to include prolongations, word repetitions, sound repetitions, blocks, and interjections (including filler words or hesitation sounds).

\paragraph{Speech Expressiveness} The last dynamic trait we aim to model is speech expressiveness. Similar to voice quality, the datasets available are scarce in this field. We adopt the definitions provided in the \texttt{ParaSpeechCaps}~\citep{diwan2025scaling} dataset to guide our modeling. Specifically, we categorize speech expressiveness into five types: animated, laughing, passive, whispered, and enunciated.

\begin{figure} {
    \centering
    
    \includegraphics[width=\linewidth]{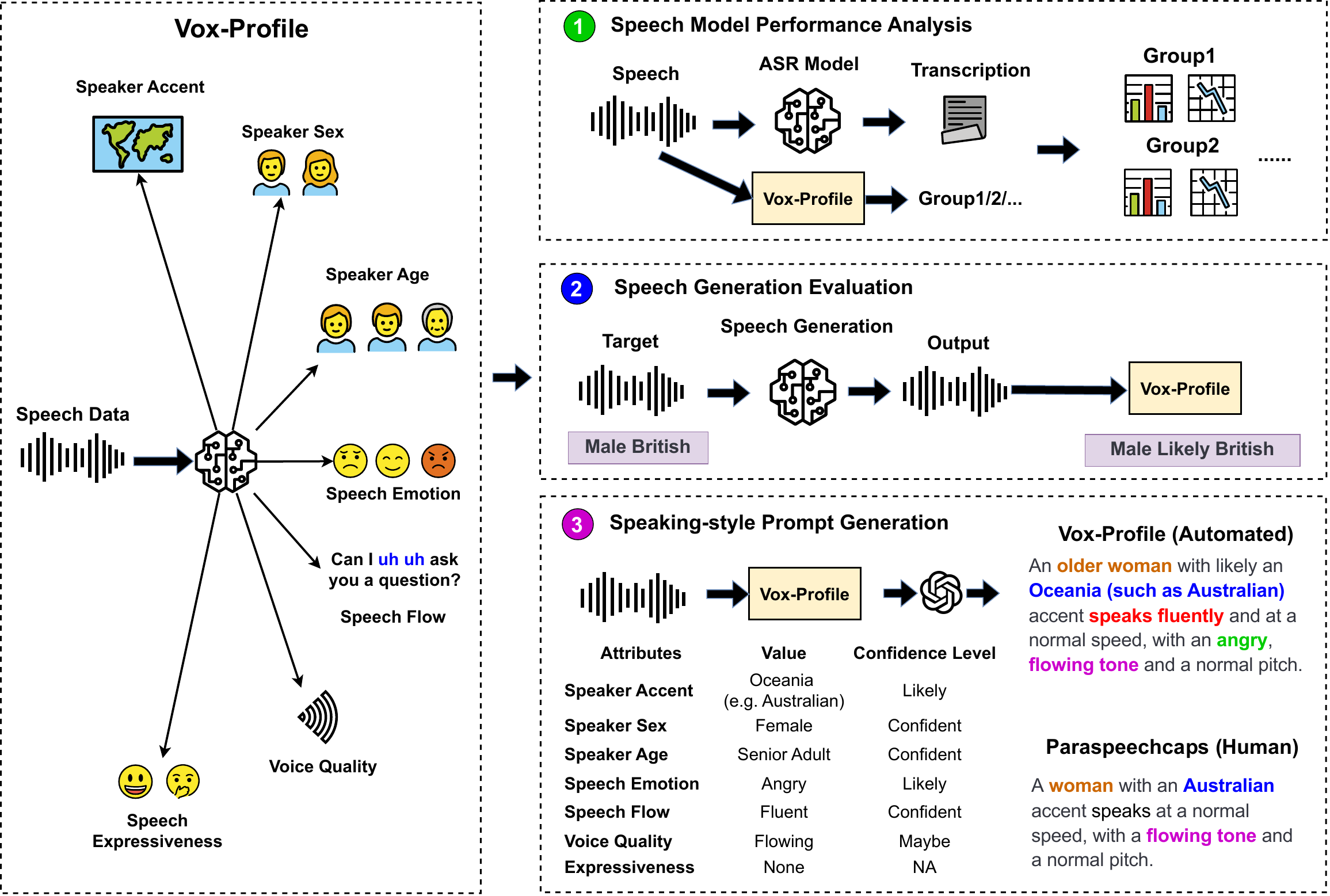}

    \caption{Overview of the \texttt{Vox-Profile} Benchmark and its applications. We highlight three primary use cases: (1) speech model (such as ASR) performance analysis, (2) automated speech generation evaluation, and (3) automated speaking style tagging. The figure uses sources from the OpenMoji web.}
    \label{fig:vox_profile_diag}
} \end{figure}

\section{Vox-Profile Benchmark}

As shown in Figure~\ref{fig:vox_profile_diag}, we propose a speaker traits and speech property modeling benchmark, \texttt{Vox-Profile}, to model a wide range of speaker traits and speech properties defined above. Our \texttt{Vox-Profile} benchmark can be applied to augment metadata for analyzing speech model performances, evaluating speech generation systems, and curating synthetic speaking style descriptions.

\subsection{Benchmark Overview}
\paragraph{Static Speaker Traits} We examine the aforementioned speech foundation models to predict different static traits from speech. (1) For age and sex classification, we experiment with the multitask learning approach to jointly predict both traits. This is motivated by the existing studies that show changes in voice characteristics across age can differ significantly between males and females. Modeling them jointly is expected to allow the system to better capture these interactions and improve classification performance. We apply the concordance correlation coefficient (CCC)~\citep{lawrence1989concordance} loss to train age prediction systems. Instead of reporting regression results from absolute age values, we map ages into the three predefined age groups and report the corresponding classification performance. (2) We develop accent classification systems in two levels of categorization: broad and narrow. The former targets broader accent categories, while the latter predicts 16 narrow accent labels based on different regional and language backgrounds. (3) We formulate voice quality prediction as a multi-label classification task since each speech sample can be associated with multiple voice qualities.

\paragraph{Dynamic Speech Traits} (1) Our categorical emotion and dimensional emotion (arousal / valence) modeling systems are built upon our recent top-ranked solutions \citep{lertpetchpun2025developing, feng2025developing} in the Speech Emotion Recognition in Naturalistic Conditions Challenge at Interspeech 2025 (IS25-SER challenge)~\footnote{\href{https://lab-msp.com/MSP-Podcast_Competition/IS2025/}{https://lab-msp.com/MSP-Podcast\_Competition/IS2025}}. Specifically, the categorical emotion system applies the KL divergence loss to predict the distribution of categorical emotion annotations. For arousal and valence modeling, we use the sigmoid function to map arousal and valence values onto a normalized scale between 0 and 1. (2) For speech flow modeling, we propose a multitask learning framework of 2 classification heads with one to predict whether the speech is fluent or disfluent, the other to classify the specific types of disfluencies present. We highlight that classifying types of disfluencies is a multilabel task, as multiple breaks can co-occur in a single speech utterance. (3) We formulate speech expressiveness as a multilabel classification task. 

\subsection{Data Preprocessing}
For all experimental datasets used in speaker traits and speech properties classification, we resample the audio to 16 kHz. We exclude audio samples shorter than 3 seconds, as reliably estimating speaker traits or speech properties is not feasible with shorter speech utterances. We also truncate the audio to 15 seconds due to the computation limitation in training. Audios with corrupted formats are removed from the experiments. For speaker accent classification, we discard samples with labels like British, as it does not specify the regional varieties.

\subsection{Modeling Approach} 
In \texttt{Vox-Profile}, we evaluate several widely studied speech foundation models, including the \texttt{HuBERT}~\citep{hsu2021hubert}, \texttt{WavLM}~\citep{hu2024wavllm}, \texttt{ECAPA-TDNN}~\citep{desplanques2020ecapa}, and \texttt{Whisper}~\citep{radford2023robust} family with details reported in Appendix~\ref{appendix-models}. Since speaker embedding models are primarily related to static speaker traits, we exclude the \texttt{ECAPA-TDNN} model for predicting dynamic speech traits. For all experiments, we use the downstream model architecture developed in~\citep{pepino2021emotion}, where many existing works have shown that this simple architecture achieves strong performance in a wide range of downstream speech modeling tasks. Specifically, we first apply weighted averaging to combine hidden representations of both the convolutional and Transformer~\citep{vaswani2017attention} encoder layers. The aggregated output is then processed through 1D-pointwise convolutional layers. Finally, we average the convolutional outputs to obtain the final embeddings, which are passed through fully connected layers for classification or regression. In addition to this approach, we experiment with finetuning using Low-Rank Adaptation, \texttt{LoRa}~\citep{hu2022lora}, an effective method for adapting speech foundation models. Details of the training hyperparameters, such as learning rates and training epochs, and training resources, are provided in Appendix~\ref{appendix-modeling} and \ref{appendix-resouce}.

\begin{table}
    \scriptsize
    \centering

    \caption{Comparison of different speech foundation models in predicting static traits. Overall, the results show that \texttt{Whisper} Large achieves the overall best performance, while the \texttt{WavLM}-based model shows moderate strength in accent prediction.}
    \vspace{1mm}
    \resizebox{0.98\linewidth}{!}{
        \begin{tabular}{lcccccccccc}
    
            \toprule
             & 
            \multicolumn{2}{c}{\textbf{Speaker Age}} & 
            \multicolumn{2}{c}{\textbf{Speaker Sex}} & 
            \multicolumn{2}{c}{\textbf{Broad Accent}} & 
            \multicolumn{2}{c}{\textbf{Narrow Accent}} & 
            \multicolumn{1}{c}{\textbf{Voice Quality}}
            \\
    
            & 
            \multicolumn{1}{c}{\textbf{Acc}} & 
            \multicolumn{1}{c}{\textbf{F1}} & 
            \multicolumn{1}{c}{\textbf{Acc}} & 
            \multicolumn{1}{c}{\textbf{F1}} & 
            \multicolumn{1}{c}{\textbf{Acc}} & 
            \multicolumn{1}{c}{\textbf{F1}} & 
            \multicolumn{1}{c}{\textbf{Acc}} & 
            \multicolumn{1}{c}{\textbf{F1}} & 
            \multicolumn{1}{c}{\textbf{F1}} 
            \\
            \cmidrule(lr){1-1} \cmidrule(lr){2-3} \cmidrule(lr){4-5} \cmidrule(lr){6-7}
            \cmidrule(lr){8-9} \cmidrule(lr){10-10}
    
            \textbf{Speaker Model} \\
            \hspace{2mm}\rotatebox[origin=c]{180}{$\Lsh$}ECAPA-TDNN & 
            59.3 & 0.469 & 
            96.9 & 0.962 & 
            74.6 & 0.638 &
            42.3 & 0.363 & 0.706
            \\
    
            \textbf{Self-Supervised} \\
            
            \hspace{2mm}\rotatebox[origin=c]{180}{$\Lsh$}HuBERT Large & 
            57.1 & 0.501 & 
            96.7 & 0.960 & 
            85.4 & 0.841 & 
            49.1 & 0.443 & 0.692
            \\
            
            \hspace{2mm}\rotatebox[origin=c]{180}{$\Lsh$}WavLM Large &
            67.6 & 0.624 & 
            97.7 & 0.971 &
            {\textbf{92.9}} & 0.884 & 
            65.7 & 0.617 & 0.703
            \\
    
            \textbf{Whisper Family} \\
            \hspace{2mm}\rotatebox[origin=c]{180}{$\Lsh$}Whisper Tiny & 
            58.1 & 0.520 & 
            94.3 & 0.934 & 
            89.1 & 0.831 & 
            54.1 & 0.484 & 0.587
            \\
    
            \hspace{2mm}\rotatebox[origin=c]{180}{$\Lsh$}Whisper Small & 
            65.8 & 0.558 & 
            96.4 & 0.958 & 
            90.9 & 0.863 &
            70.4 & 0.663 & {\textbf{0.737}}
            \\
    
            \hspace{2mm}\rotatebox[origin=c]{180}{$\Lsh$}Whisper Large & 
            {\textbf{69.5}} & {\textbf{0.643}} & 
            {\textbf{98.0}} & {\textbf{0.975}} & 
            92.8 & {\textbf{0.889}} & 
            {\textbf{75.8}} & {\textbf{0.724}} & 0.704 
            \\
    
            \bottomrule
    
        \end{tabular}
    }
    \label{tab:speaker_attribute_result}
\end{table}

\subsection{Evaluation Metrics} 

We compare different models in predicting speaker and speech traits, including biological sex, age, accent, categorical emotions, and speech fluency/disfluency, using both accuracy and macro F1 scores. Additionally, we report the macro F1 score when comparing models predicting speech voice quality, specific speech disfluencies, and speech expressiveness. Finally, we assess the predictions of arousal and valence using the concordance correlation coefficient (CCC)~\citep{lawrence1989concordance} score.

\section{Benchmark Performance}
In this section, we present a detailed comparison of benchmark performance. We report the top-performing scores for each model in Table~\ref{tab:speaker_attribute_result} and Table~\ref{tab:speech_attribute_result}. We then analyze the impact of model ensembling, followed by a comparison with existing works from the literature.

\subsection{Overall Benchmark Performance}

\paragraph{Static Speaker Traits} We compare the performance of different speech foundation models in predicting static speaker traits shown in Table~\ref{tab:speaker_attribute_result}. The results indicate a positive correlation between model parameter size and performance, with larger models generally achieving better performance in speaker trait classification. Moreover, although the \texttt{ECAPA-TDNN} model is optimized for speaker recognition tasks, finetuning \texttt{ECAPA-TDNN} still yields lower performance than self-supervised or supervised speech foundation models, particularly in accent and age classification. We observe that sex classification is relatively straightforward, with most models achieving around 95\% accuracy. Moreover, classifying between broad accent groups (North American, British Isles, and other regional or language backgrounds) yields strong performance, with macro F1 scores exceeding 0.80 in most models. However, narrow accent prediction remains challenging with only \texttt{Whisper Large} achieving a 0.724 macro F1 score in this task.

\paragraph{Dynamic Speech Traits} We present results of dynamic traits prediction in Table~\ref{tab:speech_attribute_result}. Consistent with findings for static trait prediction, we find that larger models perform better, with \texttt{Whisper Large} and \texttt{WavLM Large} achieving the highest scores across most tasks. Moreover, detecting disfluency in speech flow is relatively straightforward, with multiple models achieving over 80\% accuracy. In contrast, the benchmark results indicate that SER remains challenging given the relatively low performance across most experiments. We highlight that most models can only reach macro F1 scores around 0.4 (including our models ranked as the top-performing solutions in the IS25-SER challenge).

\begin{table}
    \footnotesize
    \centering

    \caption{Comparison of different models in predicting dynamic speech traits.}
    \vspace{1mm}
    \resizebox{\linewidth}{!}{
        \begin{tabular}{lcccccccc}
    
            \toprule
             & 
            \multicolumn{1}{c}{\textbf{Speech Emotion}} & 
            \multicolumn{1}{c}{\textbf{Arousal}} & 
            \multicolumn{1}{c}{\textbf{Valence}} & 
            \multicolumn{2}{c}{\textbf{Fluency}} & 
            \multicolumn{1}{c}{\textbf{Disfluency Type}} & 
            \multicolumn{1}{c}{\textbf{Speech Expressivenss}}
            \\
    
            & 
            \multicolumn{1}{c}{\textbf{F1}} & 
            \multicolumn{1}{c}{\textbf{CCC}} & 
            \multicolumn{1}{c}{\textbf{CCC}} & 
            \multicolumn{1}{c}{\textbf{Acc}} & 
            \multicolumn{1}{c}{\textbf{F1}} & 
            \multicolumn{1}{c}{\textbf{F1}} &
            \multicolumn{1}{c}{\textbf{F1}} 
            \\
            \cmidrule(lr){1-1} \cmidrule(lr){2-2} \cmidrule(lr){3-4} \cmidrule(lr){5-6}
            \cmidrule(lr){7-7} \cmidrule(lr){8-8}

            \textbf{Self-Supervised} \\
            \hspace{2mm}\rotatebox[origin=c]{180}{$\Lsh$}HuBERT Large & 
            0.395 & 
            0.585 & 0.650 & 
            75.5 & 0.755 & 
            0.627 & 0.565
            \\
            
            \hspace{2mm}\rotatebox[origin=c]{180}{$\Lsh$}WavLM Large &
            0.406 & 0.585& {\textbf{0.665}} &
            {\textbf{80.7}} & 
            {\textbf{0.806}} & 
            {\textbf{0.691}} & 0.689
            \\
            
            \textbf{Whisper Family} \\
            \hspace{2mm}\rotatebox[origin=c]{180}{$\Lsh$}Whisper Tiny & 
            0.337 & 0.545& 0.522& 
            76.4 & 0.762 & 
            0.639 & 0.484 
            \\
    
            \hspace{2mm}\rotatebox[origin=c]{180}{$\Lsh$}Whisper Small & 
            0.387 & 0.583
             & 0.623 & 
            80.4 & 0.804 &
            0.653 & 0.751
            \\
    
            \hspace{2mm}\rotatebox[origin=c]{180}{$\Lsh$}Whisper Large & 
            {\textbf{0.416}} & {\textbf{0.588}}
            & 0.645 & 
            80.6 & 0.805 & 
            0.670 & {\textbf{0.848}}
            \\
    
            \bottomrule
    
        \end{tabular}
    }
    \label{tab:speech_attribute_result}
\end{table}

\paragraph{Model Ensemble} We investigate whether ensembling top-performing models further improves the performance. The results in Table~\ref{tab:abalation_ensemble} show that naïve ensembles of the two top-performing models consistently yield moderate improvements compared to individual models. In particular, we observe substantial gains in modeling static traits such as speech voice quality, age, and sex.

\begin{table}
    
    \caption{Experiments to compare \texttt{Vox-Profile} with existing work and impact of model ensembles.}

    \centering

    \begin{subfigure}{0.36\linewidth}
        \vspace{-5.5em}
        \includegraphics[width=0.94\linewidth]{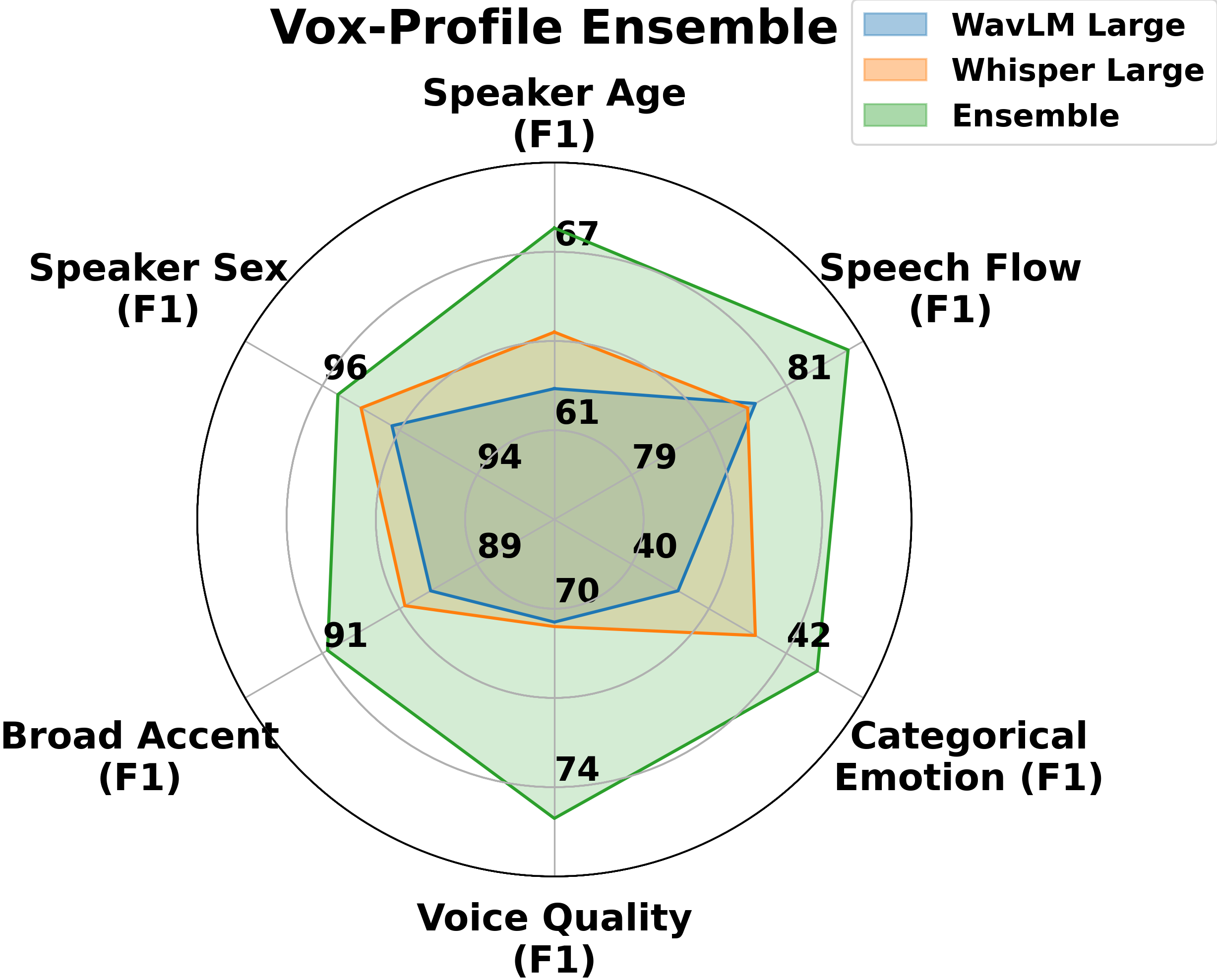}
        \caption{Comparing model ensembles with single model in \texttt{Vox-Profile}.}
        \label{tab:abalation_ensemble}
    \end{subfigure}
    \begin{subfigure}{0.01\linewidth}
    \end{subfigure}
    \begin{subtable}{0.62\linewidth}
    \resizebox{\linewidth}{!}{
        \begin{tabular}{lcclcccccc}
        \toprule 
        \textbf{Model} & \multicolumn{1}{c}{\textbf{Acc}} & \multicolumn{1}{c}{\textbf{F1}} & \textbf{Model} & \multicolumn{1}{c}{\textbf{Acc}} & \multicolumn{1}{c}{\textbf{F1}}  \\ 
        \cmidrule(lr){1-1} \cmidrule(lr){2-3}  \cmidrule(lr){4-4} \cmidrule(lr){5-6}

        \textbf{Accent (VCTK)} & & &  \textbf{Accent (British Isles)} & & \\
        \hspace{2mm}\rotatebox[origin=c]{180}{$\Lsh$}\texttt{CommonAccent} & 66.3 & 0.379 & \hspace{2mm}\rotatebox[origin=c]{180}{$\Lsh$}\texttt{CommonAccent} & 69.4 & 0.581   \\
        \hspace{2mm}\rotatebox[origin=c]{180}{$\Lsh$}\texttt{Vox-Profile} & {\textbf{72.6}} & {\textbf{0.423}} &  \hspace{2mm}\rotatebox[origin=c]{180}{$\Lsh$}\texttt{Vox-Profile} & {\textbf{83.9}} & {\textbf{0.770}}  \\
        \midrule

        \textbf{Speaker Age (VoxCeleb)} & & &  \textbf{Speaker Sex (TIMIT)} & & \\
        \hspace{2mm}\rotatebox[origin=c]{180}{$\Lsh$}\texttt{Wav2Vec2-based} & 74.9 & {\textbf{0.717}} & \hspace{2mm}\rotatebox[origin=c]{180}{$\Lsh$}\texttt{Wav2Vec2-based} & 99.4 & 0.993   \\
        \hspace{2mm}\rotatebox[origin=c]{180}{$\Lsh$}\texttt{Vox-Profile} & {\textbf{75.6}} & 0.716 &  \hspace{2mm}\rotatebox[origin=c]{180}{$\Lsh$}\texttt{Vox-Profile} & {\textbf{99.5}}  & {\textbf{0.995}} \\
        \midrule

        \textbf{Model} & \multicolumn{2}{c}{\textbf{F1}} & \textbf{Model} & \multicolumn{2}{c}{\textbf{CCC}}  \\ 
        \cmidrule(lr){1-1} \cmidrule(lr){2-3}  \cmidrule(lr){4-4} \cmidrule(lr){5-6}
        
        \textbf{Categorical Emotion} & & & \textbf{Arousal / Valence} &  &  \\
        \hspace{2mm}\rotatebox[origin=c]{180}{$\Lsh$}\texttt{IS25-SAIL SER} & \multicolumn{2}{c}{{\textbf{0.428}}} & \hspace{2mm}\rotatebox[origin=c]{180}{$\Lsh$}\texttt{IS25-SAIL SER} & \multicolumn{2}{c}{{\textbf{0.642}} / {\textbf{0.683}}} \\
        \hspace{2mm}\rotatebox[origin=c]{180}{$\Lsh$}\texttt{Vox-Profile} & \multicolumn{2}{c}{0.392} & \hspace{2mm}\rotatebox[origin=c]{180}{$\Lsh$}\texttt{Vox-Profile} & \multicolumn{2}{c}{0.623 / 0.649} \\

        \bottomrule
        \end{tabular}
    }
    \vspace{1mm}
    \caption{Comparing \texttt{Vox-Profile} benchmark performance with the existing literature.}
    \label{tab:abalation_existing}
    \end{subtable}

    \label{tab:abalation}
\end{table}

\subsection{Comparing Vox-Profile Benchmark with Existing Literature} 

We compare the benchmark performance of \texttt{Vox-Profile} to several prior works. Regarding the accent classification, we compare our model against a previous approach CommonAccent~\citep{zuluaga2023commonaccent} on the VCTK~\citep{vctk} and British Isles~\citep{demirsahin-etal-2020-open} speaker datasets. To ensure a fair comparison, we mapped the label schema used in CommonAccent to align with our accent labeling. As shown in Table~\ref{tab:abalation_existing}, our benchmark model consistently outperforms CommonAccent in both datasets. 
Moreover, we compare our speaker age and sex classification model with the approach presented in \citep{burkhardt2023speech}, using the test sets of the VoxCeleb and TIMIT datasets, respectively. The results indicate that \texttt{Vox-Profile} achieves strong performance in both age and sex prediction tasks.
Finally, we compare our SER models with our top-performing system, \texttt{SAIL-SER}, which achieves top-1 and top-2 scores in the categorical and dimensional SER in the IS25-SER challenge. Note that \texttt{SAIL-SER} utilizes both text and speech modalities, while \texttt{Vox-Profile} relies solely on the speech input. 
The absence of text modality in \texttt{Vox-Profile} explains the performance gap; however, it still achieves comparable performance despite this limitation.

\section{Enabling Versatile Speech Applications with Vox-Profile}

\subsection{Speech Model Performance Analysis}

We show how \texttt{Vox-Profile} facilitates the analysis of speech model performance. Specifically, we generate speaker and speech traits for existing datasets and investigate whether these generated labels can lead to the same insights as using the ground truth trait information in analyzing the speech model performances. For this experiment, we focus on ASR as a representative task in speech modeling.

\paragraph{ASR Inference} We augment accent and emotion labels for the VCTK and MSP-Podcast datasets. For MSP-Podcast, we provide emotion labels only for the dev set, following the 4 most frequently represented emotions: neutral, sadness, happiness, and anger. Only samples with confident predictions are used for computing ASR performance in the \texttt{Vox-Profile} experiment.
To mitigate the potential artifacts from using the same model architecture for ASR and speech trait prediction, we use the \texttt{Wav2Vec 2.0 Robust}~\citep{hsu2021robust} to infer the transcript. In our analysis, we exclude the ASR models, which the VCTK dataset is trained on, such as \texttt{Parakeet}~\citep{rekesh2023fast} or \texttt{Canary}~\citep{rekesh2023fast}, for fair comparison.

\begin{figure} {
    \centering
    \begin{tikzpicture}

        \node[draw=none,fill=none] at (0,0){\includegraphics[width=0.5\linewidth]{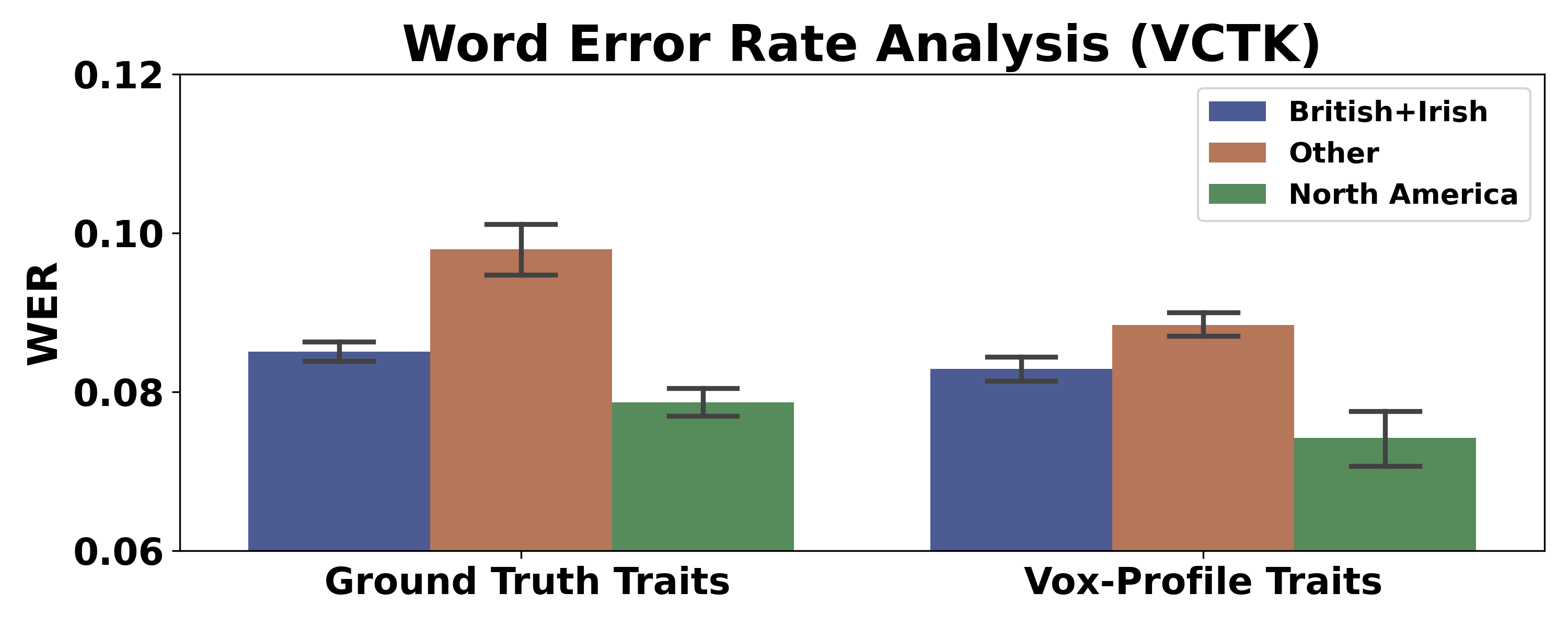}};

        \node[draw=none,fill=none] at (0.5\linewidth,0){\includegraphics[width=0.5\linewidth]{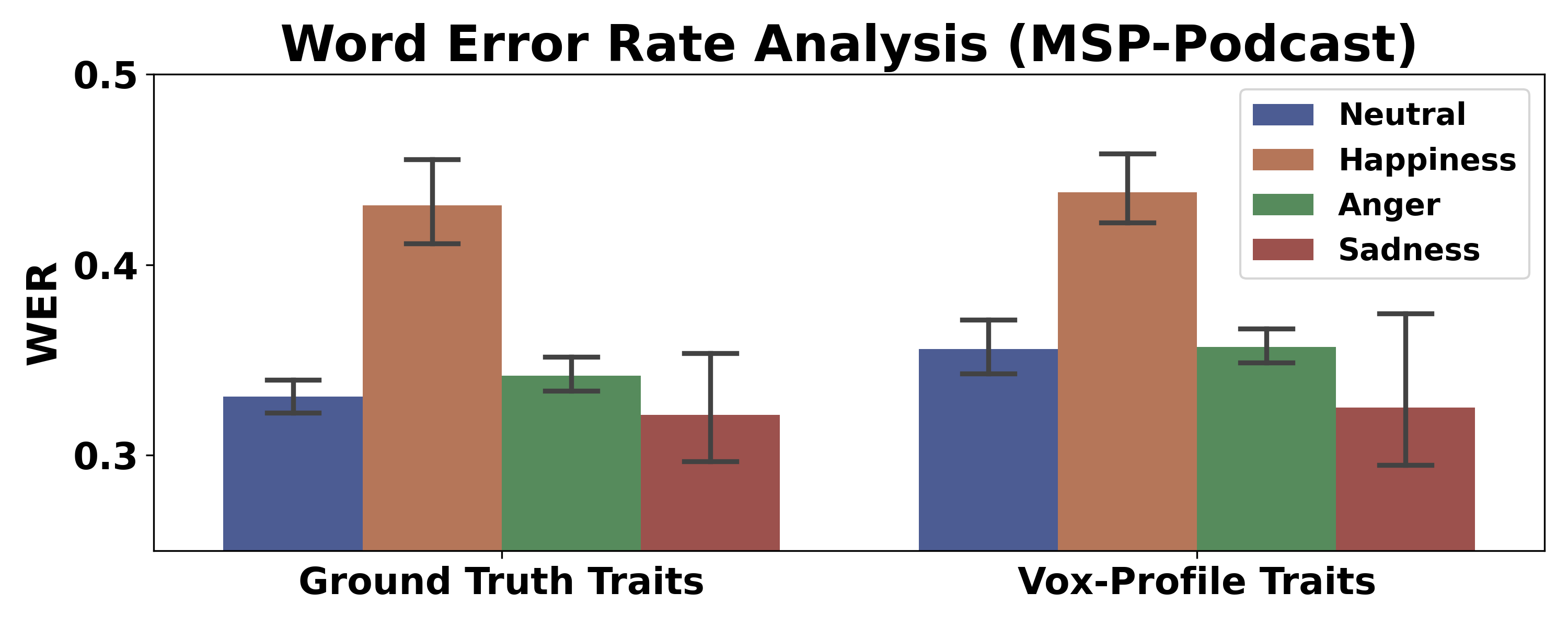}};
        
    \end{tikzpicture}
    \caption{ASR performance trends, grouped by ground truth and predicted labels by \texttt{Vox-Profile}. We measure WER, stratified by accent and emotion labels. We observe similar performance trends between the predicted and ground truth trait labels.}
    \label{fig:vox_profile_asr}
} \end{figure}

\paragraph{Findings} Figure~\ref{fig:vox_profile_asr} shows that ASR performance trends based on the ground truth trait align closely with those using the predicted labels from \texttt{Vox-Profile}. In the VCTK dataset, North American speakers consistently have lower WER scores than speakers from the British Isles and those with ``Other'' accents, regardless of whether ground truth or synthetic labels are used ($p<0.01$ in both conditions). Moreover, whether using ground truth or predicted emotion labels, in this particular dataset, speech expressing sadness is associated with relatively lower WER scores, whereas happy speech tends to result in higher WER scores. Overall, these findings suggest that \texttt{Vox-Profile} can generate reliable synthetic metadata to facilitate in-depth analysis of speech model performance.

\subsection{Automated Evaluation Tool for Speech Generation Tasks}

We demonstrate the utility of \texttt{Vox-Profile} as an evaluation tool for speech generation tasks by comparing two representative models: \texttt{FreeVC}~\citep{freevc} and \texttt{VALLE-X}~\citep{vallex}.
\texttt{FreeVC} is selected as a representative of textless voice conversion models that operate in the latent space, aiming to transform source speech to match the voice identity of a reference speaker. \texttt{VALLE-X} is chosen as a representative of voice cloning approaches based on neural codecs, utilizing a concatenated pipeline of ASR and TTS conditioned on the reference speech.
It has been reported that textless models like \texttt{FreeVC} often struggle to accurately reflect the accent of the reference speaker~\citep{baade24_interspeech}. On the other hand, \texttt{VALLE-X}, by incorporating an intermediate text representation through ASR, is better equipped to preserve accentual features in the generated speech.

To assess the ability of these models to reflect the accent of the reference speech, we randomly selected source-reference (target) pairs from the VCTK dataset~\citep{vctk}. For each pair, we synthesized speech samples and evaluated whether they more closely resembled the source or reference speaker in terms of accent, by measuring cosine similarities and accent prediction scores. Detailed configurations for \texttt{FreeVC}, \texttt{VALLE-X}, and the selected test speakers are provided in Appendix~\ref{appendix-speechgentask}.

As shown in Table~\ref{tab:voice_cloning_similarity}, the accent prediction scores and the cosine similarity for the synthesized samples from \texttt{FreeVC} suggest greater similarity to the source speaker’s accent than to the reference speaker. In contrast, the scores for \texttt{VALLE-X} indicate closer alignment with the reference speaker’s accent in most conditions. These findings are consistent with previous studies, which report that \texttt{FreeVC} has limited capability in replicating the accentual features of the reference speaker, whereas \texttt{VALLE-X}, due to its intermediate text-based representation, more effectively preserves these features~\citep{baade24_interspeech, vallex}.

\subsection{Generating Synthetic Speaking Style Prompt}

We further apply the \texttt{Vox-Profile} as a tool for generating synthetic speaker and speech traits to create speaking style prompts. Previous efforts to create synthetic speaking style prompts, such as \texttt{ParlerTTS}~\citep{lyth2024natural}, rely on heuristic-based categorizations of acoustic features, including fundamental frequency (F0) and signal-to-noise ratio (SNR). More recently, \texttt{ParaSpeechCaps} introduced a large-scale, human-annotated dataset containing multiple speech traits, such as voice quality and speaker accent. In contrast to these prior works, \texttt{Vox-Profile} provides a more extensive and varied set of traits, including speech flow, arousal, valence, and speaker age. Moreover, computational models of \texttt{Vox-Profile} output probabilistic predictions for each trait, enabling more nuanced and confidence-sensitive descriptions. For example, a Scottish accent prediction with a probability of 0.9 can be described as having a distinct Scottish accent, while a probability of 0.5 might be described as a likely Scottish accent. This probabilistic framing supports the generation of speech descriptions that better reflect the para-linguistic uncertainties with which humans naturally perceive and interpret speech for information extraneous to linguistic messaging. To evaluate the robustness and versatility of the style prompts generated by \texttt{Vox-Profile}, we propose a comparative analysis between human-annotated and machine-generated style prompts using the \texttt{ParaSpeechCaps} dataset. The detailed human evaluation procedure is described in Appendix~\ref{appendix-textdescripteval}.

\begin{table}[t]

\caption{Utilizing \texttt{Vox-Profile} to evaluate the accent conversion performance of \texttt{FreeVC} and \texttt{VALLE-X}, by measuring cosine similarity (left table) and prediction scores (right table).}

\begin{minipage}[t]{0.62\textwidth}
\resizebox{\textwidth}{!}{
    \begin{tabular}{lcccccccc}

        \toprule
         & \multicolumn{2}{c}{\textbf{Similarity Between}} & 
        \\

        \textbf{Source $\Rightarrow$ Reference} & \multicolumn{1}{c}{\textbf{(Source, Output) $\downarrow$}} & 
        \multicolumn{1}{c}{\textbf{(Reference, Output) $\uparrow$}}
        \\

        \cmidrule(lr){1-1} \cmidrule(lr){2-3}

        \textbf{British Isles $\Rightarrow$ North America} \\
        \hspace{2mm}\rotatebox[origin=c]{180}{$\Lsh$}FreeVC & 
        \textcolor{purple}{\textbf{0.874 $\pm$ 0.046}} & 0.319 $\pm$ 0.071
        \\

        \hspace{2mm}\rotatebox[origin=c]{180}{$\Lsh$}VALLE-X &
        0.512 $\pm$ 0.103 & \textcolor{teal}{\textbf{0.543 $\pm$ 0.080}}
        \\
        
        \textbf{British Isles $\Rightarrow$ Other} \\
        \hspace{2mm}\rotatebox[origin=c]{180}{$\Lsh$}FreeVC & 
        \textcolor{purple}{\textbf{0.818 $\pm$ 0.064}} & 0.552 $\pm$ 0.089
        \\

        \hspace{2mm}\rotatebox[origin=c]{180}{$\Lsh$}VALLE-X &
        0.497 $\pm$ 0.102 & \textcolor{teal}{\textbf{0.908 $\pm$ 0.055}}
        \\

        \midrule
        \textbf{North America $\Rightarrow$ British Isles} \\
        \hspace{2mm}\rotatebox[origin=c]{180}{$\Lsh$}FreeVC & 
        \textcolor{purple}{\textbf{0.846 $\pm$ 0.046}} & 0.503 $\pm$ 0.086
        \\
        
        \hspace{2mm}\rotatebox[origin=c]{180}{$\Lsh$}VALLE-X &
        0.559 $\pm$ 0.082 & \textcolor{teal}{\textbf{0.595 $\pm$ 0.090}}
        \\

        \textbf{Other $\Rightarrow$ British Isles} \\
        \hspace{2mm}\rotatebox[origin=c]{180}{$\Lsh$}FreeVC & 
        {0.937 $\pm$ 0.036} & \textcolor{teal}{\textbf{0.658 $\pm$ 0.090}}
        \\
        
        \hspace{2mm}\rotatebox[origin=c]{180}{$\Lsh$}VALLE-X &
        \textcolor{purple}{\textbf{0.960 $\pm$ 0.030}} & 0.552 $\pm$ 0.097
        \\
        
        \bottomrule

    \end{tabular}
}
\end{minipage}
\begin{minipage}{0.37\textwidth}
\centering
\includegraphics[width=0.9\linewidth]{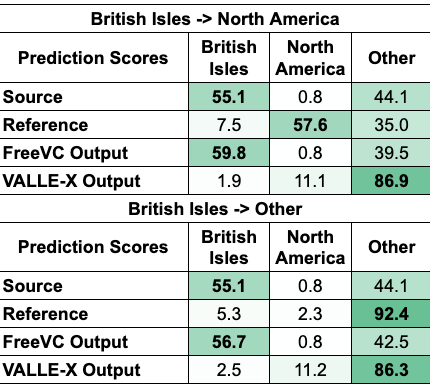}
\end{minipage}
\label{tab:voice_cloning_similarity}
\end{table}

\begin{figure} {
    \centering
    \includegraphics[width=\linewidth]{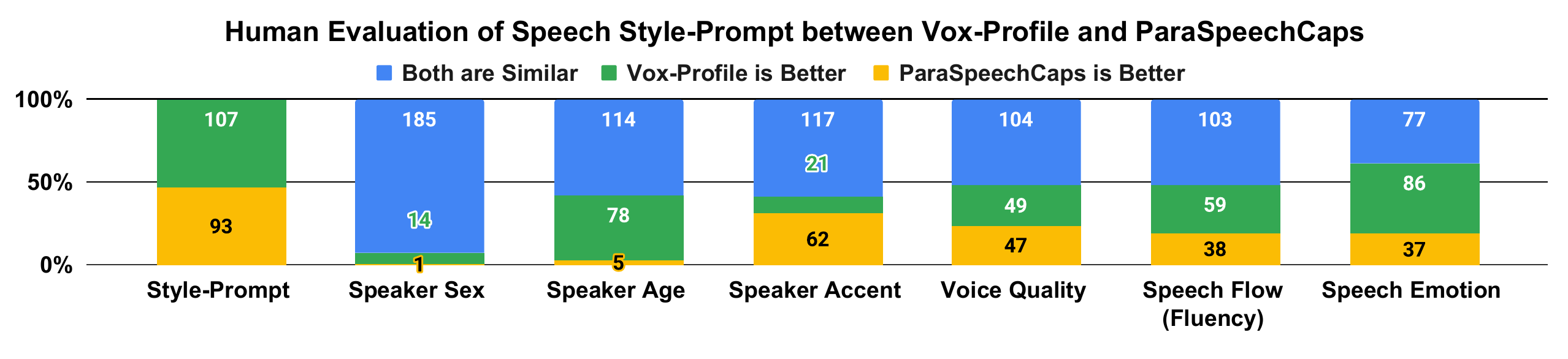}
    \caption{Comparing human evaluation results between synthetic speaking style prompts created by \texttt{Vox-Profile} and human-annotated prompts from \texttt{ParaSpeechCaps}. Human raters provided their preferences across overall prompt quality, sex, age, accent, voice quality, fluency, and emotion.}
    \label{fig:human_eval_results}
} \end{figure}

\paragraph{Prompt Generation} Since our voice quality and expressiveness models are trained on the training set of the \texttt{ParaSpeechCaps}, we use speech samples from the test set for this experiment. Specifically, we select 30 speech samples stratified by speaker sex and accent (broader accent classes). For each selected sample, we use \texttt{Vox-Profile} to infer both static and dynamic traits. To ensure reliability, we keep the predictions with high and moderate confidence levels, labeling them as "distinct/very likely" and "likely," respectively. Traits with low confidence are excluded from prompt generation. Subsequently, we use \texttt{GPT-4.1}~\citep{openai2024gpt4} to generate the speaking style prompt using the synthetic tags. To ensure a fair comparison, we also apply the \texttt{GPT-4.1} to generate the speaking style prompt using the \texttt{ParaSpeechCaps} tags. The prompt generation follows the template used in \texttt{ParaSpeechCaps} and the detailed description of confidence levels and prompt designs is in Appendix~\ref{appendix-confidence}.

\paragraph{Human-Evaluation Results} Human-evaluation results comparing synthetic speaking style prompts from \texttt{Vox-Profile} and human-annotated speaking style prompts from Para-SpeechCaps are presented in Figure~\ref{fig:human_eval_results}. Overall, the results suggest that this group of human raters shows similar preference levels for both synthetic and human-annotated speaking style prompts. Specifically, they favor the emotion, age, and speech flow descriptions generated by \texttt{Vox-Profile} over those from \texttt{ParaSpeechCaps}. For accent descriptions, the human raters reach similar levels of agreement between \texttt{ParaSpeechCaps} and \texttt{Vox-Profile} in more than 50\% unique evaluations. However, ground truth accent labels from \texttt{ParaSpeechCaps} still outperform \texttt{Vox-Profile} in accurately describing speaker accents, highlighting that accent prediction remains challenging in speaker trait modeling. In summary, the human evaluation results provide evidence suggesting that \texttt{Vox-Profile} can be used to create synthetic speaking style prompts that closely match human-labeled data.

\section{Ethical Use and Mitigation of Misuses}

It is critical to recognize that \texttt{Vox-Profile} predicts speaker group-level traits and does not perform speaker recognition. Privacy risks are therefore different but not absent. The intended use of \texttt{Vox-Profile} includes analysis, evaluation, and dataset augmentation, and it should not be used for individual profiling or eligibility decisions. While speaker recognition is used to uniquely identify individuals from speech, raising direct concerns about surveillance or re-identification, Vox-Profile instead infers generalized traits such as age group, which are less personally identifiable. Nonetheless, we acknowledge that under different conditions, even these trait predictions may pose privacy risks depending on legal definitions, usage agreements, or industry practices. To mitigate potential misuses, we discuss two complementary prevention strategies below.

\paragraph{Legal Prevention} Most \texttt{Vox-Profile} models are released under a Responsible AI License (RAIL) that explicitly prohibits applications involving surveillance, discrimination, or unauthorized inference of sensitive personal data under the scope of law and regulation. For example, the license prohibits using the system to ``detect or infer any legally protected class or aspect of any person'' or to ``diagnose a medical condition without human oversight'' Users should respect the privacy and consent of the data subjects and adhere to the relevant laws and regulations in their jurisdictions when using \texttt{Vox-Profile}. Specifically, models derived from data hosted using FluencyBank~\citep{ratner2018fluency}, part of the TalkBank system~\citep{macwhinney2007talkbank}, are subject to TalkBank's Ground Rules and distributed under CC BY-NC-SA 3.0. Moreover, we encourage users to apply \texttt{Vox-Profile} only for robustness analysis and evaluation of speech technologies and not for individual re-identification. The allowed applications include evaluating limitations in ASR models or improving speech generation systems, as shown in the paper. Relevant ethical considerations include, but are not limited to, those outlined by Gao et al.~\citep{gao2024documenting}.  

\paragraph{Software} Users may consider voice anonymization techniques before applying \texttt{Vox-Profile} to prevent revealing sensitive data. For example, the researchers may consider adding noise to specific features, as shown by \citep{mireshghallah2021not}, to anonymize speaker age, but not speaker accent, when evaluating ASR models for accented voices, when consent allows such use. Users may refer to more voice anonymization techniques described in \citep{tomashenko2024voiceprivacy}.

\section{Conclusion and Limitations}
\label{sec:conclusion}

We propose \texttt{Vox-Profile}, a comprehensive benchmark for modeling speaker and speech characteristics. 
This benchmark includes large-scale experiments across more than 15 public speech datasets, evaluated using widely adopted speech foundation models, and delivers a suite of high-performing trait prediction models. 
Our results show that \texttt{Vox-Profile} supports a broad range of applications, including speech model performance analysis, generated speech evaluation, and high-quality speaker and style prompt generation. 
Our next steps include testing our benchmark in multilingual conditions and adding the prediction of multilingual speech attributes like language identification and code-switching to improve accessibility for more diverse language communities.
Moreover, while \texttt{Vox-Profile} currently uses speech models pre-trained mostly for ASR, we plan to expand to alternative architectures such as \texttt{Emotion2Vec}~\citep{ma2024emotion2vec} and \texttt{Speech-LLMs}~\citep{fathullah2024audiochatllama, hu2024wavllm}, which we expect will broaden the benchmark's applicability to affective and conversational speech technologies.

\vspace{1mm}
\noindent \textbf{Limitations} Firstly, the human evaluation in Section 6.3 is a small-scale study involving approximately 10 raters and 30 speech samples. Because the raters consist primarily of male, non-native English speakers (as shown in Appendix~\ref{appendix-textdescripteval}), the current findings may not generalize to broader listener populations. Specifically, a larger study with a more demographically balanced group of raters is needed to evaluate whether perceptions of prompt quality differ across listener groups. One other limitation is that we have not systematically verified whether our evaluation datasets overlap with the data used to pre-train the speech foundation models. For example, some Vox-Profile datasets, including Common Voice and VoxCeleb, may appear in the published pretraining or fine-tuning corpora of related speech models. Specifically, we cannot exclude potential overlap with the data used for the Whisper checkpoints.

\section{Acknowledgement}
This work was supported by the Office of the Director of National Intelligence (ODNI), Intelligence Advanced Research Projects Activity (IARPA), via the ARTS Program under contract D2023-2308110001. The views and conclusions contained herein are those of the authors and should not be interpreted as necessarily representing the official policies, either expressed or implied, of ODNI, IARPA, or the U.S. Government. The U.S. Government is authorized to reproduce and distribute reprints for governmental purposes notwithstanding any copyright annotation therein.

\bibliography{dmlr}

@inproceedings{burkhardt2023speech,
  title={Speech-based age and gender prediction with transformers},
  author={Burkhardt, Felix and Wagner, Johannes and Wierstorf, Hagen and Eyben, Florian and Schuller, Bj{\"o}rn},
  booktitle={Speech Communication; 15th ITG Conference},
  pages={46--50},
  year={2023},
  organization={VDE}
}

@inproceedings{yang2025demographic,
  title={Demographic Attributes Prediction from Speech Using WavLM Embeddings},
  author={Yang, Yuchen and Thebaud, Thomas and Dehak, Najim},
  booktitle={2025 59th Annual Conference on Information Sciences and Systems (CISS)},
  pages={1--6},
  year={2025},
  organization={IEEE}
}

@inproceedings{radford2023robust,
  title={Robust speech recognition via large-scale weak supervision},
  author={Radford, Alec and Kim, Jong Wook and Xu, Tao and Brockman, Greg and McLeavey, Christine and Sutskever, Ilya},
  booktitle={International Conference on Machine Learning},
  pages={28492--28518},
  year={2023},
  organization={PMLR}
}

@article{kuppens2013relation,
  title={The relation between valence and arousal in subjective experience.},
  author={Kuppens, Peter and Tuerlinckx, Francis and Russell, James A and Barrett, Lisa Feldman},
  journal={Psychological bulletin},
  volume={139},
  number={4},
  pages={917},
  year={2013},
  publisher={American Psychological Association}
}

@inproceedings{snyder2018x,
  title={X-vectors: Robust dnn embeddings for speaker recognition},
  author={Snyder, David and Garcia-Romero, Daniel and Sell, Gregory and Povey, Daniel and Khudanpur, Sanjeev},
  booktitle={2018 IEEE international conference on acoustics, speech and signal processing (ICASSP)},
  pages={5329--5333},
  year={2018},
  organization={IEEE}
}

@article{lotfian2017building,
  title={Building naturalistic emotionally balanced speech corpus by retrieving emotional speech from existing podcast recordings},
  author={Lotfian, Reza and Busso, Carlos},
  journal={IEEE Transactions on Affective Computing},
  volume={10},
  number={4},
  pages={471--483},
  year={2017},
  publisher={IEEE}
}

@article{hu2024wavllm,
  title={Wavllm: Towards robust and adaptive speech large language model},
  author={Hu, Shujie and Zhou, Long and Liu, Shujie and Chen, Sanyuan and Meng, Lingwei and Hao, Hongkun and Pan, Jing and Liu, Xunying and Li, Jinyu and Sivasankaran, Sunit and others},
  journal={arXiv preprint arXiv:2404.00656},
  year={2024}
}

@inproceedings{zhong2025accentbox,
  title={AccentBox: Towards High-Fidelity Zero-Shot Accent Generation},
  author={Zhong, Jinzuomu and Richmond, Korin and Su, Zhiba and Sun, Siqi},
  booktitle={ICASSP 2025-2025 IEEE International Conference on Acoustics, Speech and Signal Processing (ICASSP)},
  pages={1--5},
  year={2025},
  organization={IEEE}
}

@article{ratner2018fluency,
  title={Fluency Bank: A new resource for fluency research and practice},
  author={Ratner, Nan Bernstein and MacWhinney, Brian},
  journal={Journal of fluency disorders},
  volume={56},
  pages={69--80},
  year={2018},
  publisher={Elsevier}
}

@incollection{macwhinney2007talkbank,
  title={The talkbank project},
  author={MacWhinney, Brian},
  booktitle={Creating and digitizing language corpora: Volume 1: Synchronic databases},
  pages={163--180},
  publisher={Springer}
}

@inproceedings{lea2021sep,
  title={Sep-28k: A dataset for stuttering event detection from podcasts with people who stutter},
  author={Lea, Colin and Mitra, Vikramjit and Joshi, Aparna and Kajarekar, Sachin and Bigham, Jeffrey P},
  booktitle={ICASSP 2021-2021 IEEE International Conference on Acoustics, Speech and Signal Processing (ICASSP)},
  pages={6798--6802},
  year={2021},
  organization={IEEE}
}

@inproceedings{zhao2018l2,
  title={L2-ARCTIC: A Non-native English Speech Corpus},
  author={Zhao, Guanlong and Sonsaat, Sinem and Silpachai, Alif and Lucic, Ivana and Chukharev-Hudilainen, Evgeny and Levis, John and Gutierrez-Osuna, Ricardo},
  booktitle={Proc. Interspeech 2018},
  pages={2783--2787},
  year={2018}
}

@article{garofolo1993darpa,
  title={DARPA TIMIT acoustic-phonetic continous speech corpus CD-ROM. NIST speech disc 1-1.1},
  author={Garofolo, John S and Lamel, Lori F and Fisher, William M and Fiscus, Jonathan G and Pallett, David S},
  journal={NASA STI/Recon technical report n},
  volume={93},
  pages={27403},
  year={1993}
}

@article{nigerian_eng,
    title = {Crowdsourced high-quality Nigerian English speech data set.},
    journal = {Open Speech and Language Resources},
    author = {},
    year = {2019}
}

@article{hispanic_eng,
    author = {William Byrne and Eva Knodt and Jared Bernstein and Farzhad Emami},
    title = {Hispanic-English Database (LDC2014S05)},
    journal = {Linguistic Data Consortium},
    year = {2014}
}

@inproceedings{demirsahin2020open,
  title={Open-source multi-speaker corpora of the English accents in the British isles},
  author={Demirsahin, Isin and Kjartansson, Oddur and Gutkin, Alexander and Rivera, Clara},
  booktitle={Proceedings of the twelfth language resources and evaluation conference},
  pages={6532--6541},
  year={2020}
}

@inproceedings{fathullah2024audiochatllama,
  title={AudioChatLlama: Towards General-Purpose Speech Abilities for LLMs},
  author={Fathullah, Yassir and Wu, Chunyang and Lakomkin, Egor and Li, Ke and Jia, Junteng and Shangguan, Yuan and Mahadeokar, Jay and Kalinli, Ozlem and Fuegen, Christian and Seltzer, Mike},
  booktitle={Proceedings of the 2024 Conference of the North American Chapter of the Association for Computational Linguistics: Human Language Technologies (Volume 1: Long Papers)},
  pages={5522--5532},
  year={2024}
}

@article{michelsanti2021overview,
  title={An Overview of Deep-learning-based Audio-visual Speech Enhancement and Separation},
  author={Michelsanti, Daniel and Tan, Zheng-Hua and Zhang, Shi-Xiong and Xu, Yong and Yu, Meng and Yu, Dong and Jensen, Jesper},
  journal={IEEE/ACM Transactions on Audio, Speech, and Language Processing},
  volume={29},
  pages={1368--1396},
  year={2021},
  publisher={IEEE}
}

@article{park2022review,
  title={A review of speaker diarization: Recent advances with deep learning},
  author={Park, Tae Jin and Kanda, Naoyuki and Dimitriadis, Dimitrios and Han, Kyu J and Watanabe, Shinji and Narayanan, Shrikanth},
  journal={Computer Speech \& Language},
  volume={72},
  pages={101317},
  year={2022},
  publisher={Elsevier}
}

@article{prabhavalkar2023end,
  title={End-to-end speech recognition: A survey},
  author={Prabhavalkar, Rohit and Hori, Takaaki and Sainath, Tara N and Schl{\"u}ter, Ralf and Watanabe, Shinji},
  journal={IEEE/ACM Transactions on Audio, Speech, and Language Processing},
  volume={32},
  pages={325--351},
  year={2023},
  publisher={IEEE}
}

@inproceedings{ma2024emotion2vec,
  title={emotion2vec: Self-Supervised Pre-Training for Speech Emotion Representation},
  author={Ma, Ziyang and Zheng, Zhisheng and Ye, Jiaxin and Li, Jinchao and Gao, Zhifu and Zhang, ShiLiang and Chen, Xie},
  booktitle={Findings of the Association for Computational Linguistics ACL 2024},
  pages={15747--15760},
  year={2024}
}

@article{lyth2024natural,
  title={Natural language guidance of high-fidelity text-to-speech with synthetic annotations},
  author={Lyth, Dan and King, Simon},
  journal={arXiv preprint arXiv:2402.01912},
  year={2024}
}

@inproceedings{veliche2024towards,
  title={Towards measuring fairness in speech recognition: Fair-Speech dataset},
  author={Veliche, Irina-Elena and Huang, Zhuangqun and Ayyat Kochaniyan, Vineeth and Peng, Fuchun and Kalinli, Ozlem and Seltzer, Michael L},
  booktitle={Proc. Interspeech 2024},
  pages={1385--1389},
  year={2024}
}

@article{hu2022lora,
  title={Lora: Low-rank adaptation of large language models.},
  author={Hu, Edward J and Shen, Yelong and Wallis, Phillip and Allen-Zhu, Zeyuan and Li, Yuanzhi and Wang, Shean and Wang, Lu and Chen, Weizhu and others},
  journal={ICLR},
  volume={1},
  number={2},
  pages={3},
  year={2022}
}

@inproceedings{pepino2021emotion,
  title={Emotion Recognition from Speech Using wav2vec 2.0 Embeddings},
  author={Pepino, Leonardo and Riera, Pablo and Ferrer, Luciana},
  booktitle={Proc. Interspeech 2021},
  pages={3400--3404},
  year={2021}
}

@inproceedings{yang2021superb,
  title={SUPERB: Speech Processing Universal PERformance Benchmark},
  author={Yang, Shu-wen and Chi, Po-Han and Chuang, Yung-Sung and Lai, Cheng-I Jeff and Lakhotia, Kushal and Lin, Yist Y and Liu, Andy T and Shi, Jiatong and Chang, Xuankai and Lin, Guan-Ting and others},
  booktitle={Proc. Interspeech 2021},
  pages={1194--1198},
  year={2021}
}

@article{baevski2020wav2vec,
  title={wav2vec 2.0: A framework for self-supervised learning of speech representations},
  author={Baevski, Alexei and Zhou, Yuhao and Mohamed, Abdelrahman and Auli, Michael},
  journal={Advances in neural information processing systems},
  volume={33},
  pages={12449--12460},
  year={2020}
}

@inproceedings{nguyen2023expresso,
  title={Expresso: A Benchmark and Analysis of Discrete Expressive Speech Resynthesis},
  author={Nguyen, Tu Anh and Hsu, Wei-Ning and D'Avirro, Antony and Shi, Bowen and Gat, Itai and Fazel-Zarani, Maryam and Remez, Tal and Copet, Jade and Synnaeve, Gabriel and Hassid, Michael and others},
  booktitle={Proc. Interspeech 2023},
  pages={4823--4827},
  year={2023}
}

@inproceedings{richter2024ears,
  title={EARS: An Anechoic Fullband Speech Dataset Benchmarked for Speech Enhancement and Dereverberation},
  author={Richter, Julius and Wu, Yi-Chiao and Krenn, Steven and Welker, Simon and Lay, Bunlong and Watanabe, Shinji and Richard, Alexander and Gerkmann, Timo},
  booktitle={Proc. Interspeech 2024},
  pages={4873--4877},
  year={2024}
}

@article{wang2024usat,
  title={Usat: A universal speaker-adaptive text-to-speech approach},
  author={Wang, Wenbin and Song, Yang and Jha, Sanjay},
  journal={IEEE/ACM Transactions on Audio, Speech, and Language Processing},
  year={2024},
  publisher={IEEE}
}

@article{busso2008iemocap,
  title={IEMOCAP: Interactive emotional dyadic motion capture database},
  author={Busso, Carlos and Bulut, Murtaza and Lee, Chi-Chun and Kazemzadeh, Abe and Mower, Emily and Kim, Samuel and Chang, Jeannette N and Lee, Sungbok and Narayanan, Shrikanth S},
  journal={Language resources and evaluation},
  volume={42},
  pages={335--359},
  year={2008},
  publisher={Springer}
}

@inproceedings{wang2021voxpopuli,
  title={VoxPopuli: A Large-Scale Multilingual Speech Corpus for Representation Learning, Semi-Supervised Learning and Interpretation},
  author={Wang, Changhan and Riviere, Morgane and Lee, Ann and Wu, Anne and Talnikar, Chaitanya and Haziza, Daniel and Williamson, Mary and Pino, Juan and Dupoux, Emmanuel},
  booktitle={Proceedings of the 59th Annual Meeting of the Association for Computational Linguistics and the 11th International Joint Conference on Natural Language Processing (Volume 1: Long Papers)},
  pages={993--1003},
  year={2021}
}

@inproceedings{feng2023peft,
  title={Peft-ser: On the use of parameter efficient transfer learning approaches for speech emotion recognition using pre-trained speech models},
  author={Feng, Tiantian and Narayanan, Shrikanth},
  booktitle={2023 11th International Conference on Affective Computing and Intelligent Interaction (ACII)},
  pages={1--8},
  year={2023},
  organization={IEEE}
}

@article{wagner2023dawn,
  title={Dawn of the transformer era in speech emotion recognition: closing the valence gap},
  author={Wagner, Johannes and Triantafyllopoulos, Andreas and Wierstorf, Hagen and Schmitt, Maximilian and Burkhardt, Felix and Eyben, Florian and Schuller, Bj{\"o}rn W},
  journal={IEEE Transactions on Pattern Analysis and Machine Intelligence},
  volume={45},
  number={9},
  pages={10745--10759},
  year={2023},
  publisher={IEEE}
}

@inproceedings{zadeh2018multimodal,
  title={Multimodal language analysis in the wild: Cmu-mosei dataset and interpretable dynamic fusion graph},
  author={Zadeh, AmirAli Bagher and Liang, Paul Pu and Poria, Soujanya and Cambria, Erik and Morency, Louis-Philippe},
  booktitle={Proceedings of the 56th Annual Meeting of the Association for Computational Linguistics (Volume 1: Long Papers)},
  pages={2236--2246},
  year={2018}
}

@article{cao2014crema,
  title={Crema-d: Crowd-sourced emotional multimodal actors dataset},
  author={Cao, Houwei and Cooper, David G and Keutmann, Michael K and Gur, Ruben C and Nenkova, Ani and Verma, Ragini},
  journal={IEEE transactions on affective computing},
  volume={5},
  number={4},
  pages={377--390},
  year={2014},
  publisher={IEEE}
}

@inproceedings{poria2019meld,
  title={MELD: A Multimodal Multi-Party Dataset for Emotion Recognition in Conversations},
  author={Poria, Soujanya and Hazarika, Devamanyu and Majumder, Navonil and Naik, Gautam and Cambria, Erik and Mihalcea, Rada},
  booktitle={Proceedings of the 57th Annual Meeting of the Association for Computational Linguistics},
  pages={527--536},
  year={2019}
}

@inproceedings{sanabria2023edinburgh,
  title={The Edinburgh International Accents of english corpus: Towards the democratization of english asr},
  author={Sanabria, Ramon and Bogoychev, Nikolay and Markl, Nina and Carmantini, Andrea and Klejch, Ondrej and Bell, Peter},
  booktitle={ICASSP 2023-2023 IEEE International Conference on Acoustics, Speech and Signal Processing (ICASSP)},
  pages={1--5},
  year={2023},
  organization={IEEE}
}

@article{diwan2025scaling,
  title={Scaling Rich Style-Prompted Text-to-Speech Datasets},
  author={Diwan, Anuj and Zheng, Zhisheng and Harwath, David and Choi, Eunsol},
  journal={arXiv preprint arXiv:2503.04713},
  year={2025}
}

@article{nagrani2017voxceleb,
  title={Voxceleb: a large-scale speaker identification dataset},
  author={Nagrani, Arsha and Chung, Joon Son and Zisserman, Andrew},
  journal={arXiv preprint arXiv:1706.08612},
  year={2017}
}

@article{dehak2010front,
  title={Front-end Factor Analysis for Speaker Verification},
  author={Dehak, Najim and Kenny, Patrick J and Dehak, R{\'e}da and Dumouchel, Pierre and Ouellet, Pierre},
  journal={IEEE Transactions on Audio, Speech, and Language Processing},
  volume={19},
  number={4},
  pages={788--798},
  year={2010},
  publisher={IEEE}
}

@inproceedings{hechmi2021voxceleb,
  title={Voxceleb Enrichment for Age and Gender recognition},
  author={Hechmi, Khaled and Trong, Trung Ngo and Hautam{\"a}ki, Ville and Kinnunen, Tomi},
  booktitle={2021 IEEE Automatic Speech Recognition and Understanding Workshop (ASRU)},
  pages={687--693},
  year={2021},
  organization={IEEE}
}

@inproceedings{rekesh2023fast,
  title={Fast conformer with linearly scalable attention for efficient speech recognition},
  author={Rekesh, Dima and Koluguri, Nithin Rao and Kriman, Samuel and Majumdar, Somshubra and Noroozi, Vahid and Huang, He and Hrinchuk, Oleksii and Puvvada, Krishna and Kumar, Ankur and Balam, Jagadeesh and others},
  booktitle={2023 IEEE Automatic Speech Recognition and Understanding Workshop (ASRU)},
  pages={1--8},
  year={2023},
  organization={IEEE}
}

@inproceedings{hsu2021robust,
  title={Robust wav2vec 2.0: Analyzing Domain Shift in Self-Supervised Pre-Training},
  author={Hsu, Wei-Ning and Sriram, Anuroop and Baevski, Alexei and Likhomanenko, Tatiana and Xu, Qiantong and Pratap, Vineel and Kahn, Jacob and Lee, Ann and Collobert, Ronan and Synnaeve, Gabriel and others},
  booktitle={Proc. Interspeech 2021},
  pages={721--725},
  year={2021}
}

@article{hsu2021hubert,
  title={Hubert: Self-supervised speech representation learning by masked prediction of hidden units},
  author={Hsu, Wei-Ning and Bolte, Benjamin and Tsai, Yao-Hung Hubert and Lakhotia, Kushal and Salakhutdinov, Ruslan and Mohamed, Abdelrahman},
  journal={IEEE/ACM transactions on audio, speech, and language processing},
  volume={29},
  pages={3451--3460},
  year={2021},
  publisher={IEEE}
}

@inproceedings{desplanques2020ecapa,
  title={ECAPA-TDNN: Emphasized Channel Attention, Propagation and Aggregation in TDNN based speaker verification},
  author={Desplanques, Brecht and Thienpondt, Jenthe and Demuynck, Kris},
  booktitle={21st Annual conference of the International Speech Communication Association (INTERSPEECH 2020)},
  pages={3830--3834},
  year={2020},
  organization={International Speech Communication Association (ISCA)}
}

@article{chen2022wavlm,
  title={Wavlm: Large-scale self-supervised pre-training for full stack speech processing},
  author={Chen, Sanyuan and Wang, Chengyi and Chen, Zhengyang and Wu, Yu and Liu, Shujie and Chen, Zhuo and Li, Jinyu and Kanda, Naoyuki and Yoshioka, Takuya and Xiao, Xiong and others},
  journal={IEEE Journal of Selected Topics in Signal Processing},
  volume={16},
  number={6},
  pages={1505--1518},
  year={2022},
  publisher={IEEE}
}

@inproceedings{zuluaga2023commonaccent,
  title={CommonAccent: Exploring Large Acoustic Pretrained Models for Accent Classification Based on Common Voice},
  author={Zuluaga-Gomez, Juan and Ahmed, Sara and Visockas, Danielius and Subakan, Cem},
  booktitle={Proc. Interspeech 2023},
  pages={5291--5295},
  year={2023}
}

@inproceedings{ardila2020common,
  title={Common Voice: A Massively-Multilingual Speech Corpus},
  author={Ardila, Rosana and Branson, Megan and Davis, Kelly and Kohler, Michael and Meyer, Josh and Henretty, Michael and Morais, Reuben and Saunders, Lindsay and Tyers, Francis and Weber, Gregor},
  booktitle={Proceedings of the Twelfth Language Resources and Evaluation Conference},
  pages={4218--4222},
  year={2020}
}

@inproceedings{wang2024globe,
  title={GLOBE: A High-quality English Corpus with Global Accents for Zero-shot Speaker Adaptive Text-to-Speech},
  author={Wang, Wenbin and Song, Yang and Jha, Sanjay},
  booktitle={Proc. Interspeech 2024},
  pages={1365--1369},
  year={2024}
}

@inproceedings{baade24_interspeech,
  title     = {Neural Codec Language Models for Disentangled and Textless Voice Conversion},
  author    = {Alan Baade and Puyuan Peng and David Harwath},
  year      = {2024},
  booktitle = {Interspeech 2024},
  pages     = {182--186},
  doi       = {10.21437/Interspeech.2024-1298},
  issn      = {2958-1796},
}

@inproceedings{freevc,
  title={Freevc: Towards high-quality text-free one-shot voice conversion},
  author={Li, Jingyi and Tu, Weiping and Xiao, Li},
  booktitle={ICASSP 2023-2023 IEEE International Conference on Acoustics, Speech and Signal Processing (ICASSP)},
  pages={1--5},
  year={2023},
  organization={IEEE}
}

@article{vallex,
  title={Speak foreign languages with your own voice: Cross-lingual neural codec language modeling},
  author={Zhang, Ziqiang and Zhou, Long and Wang, Chengyi and Chen, Sanyuan and Wu, Yu and Liu, Shujie and Chen, Zhuo and Liu, Yanqing and Wang, Huaming and Li, Jinyu and others},
  journal={arXiv preprint arXiv:2303.03926},
  year={2023}
}

@misc{openai2024gpt4,
  author       = {OpenAI},
  title        = {GPT-4 Technical Report},
  year         = {2024},
  howpublished = {\url{https://openai.com/research/gpt-4}},
  note         = {Accessed: 2025-05-14}
}

@article{vctk,
    title = {CSTR VCTK Corpus: English Multi-speaker Corpus for CSTR Voice Cloning Toolkit},
    author = {Veaux, Christophe and Yamagishi, Junichi and MacDonald, Kirsten},
    journal = {University of Edinburgh. The Centre for Speech Technology Research (CSTR)},
    year = {2012}
}

@article{lawrence1989concordance,
  title={A concordance correlation coefficient to evaluate reproducibility},
  author={Lawrence, I and Lin, Kuei},
  journal={Biometrics},
  pages={255--268},
  year={1989},
  publisher={JSTOR}
}

@inproceedings{demirsahin-etal-2020-open,
    title = {{Open-source Multi-speaker Corpora of the English Accents in the British Isles}},
    author = {Demirsahin, Isin and Kjartansson, Oddur and Gutkin, Alexander and Rivera, Clara},
    booktitle = {Proceedings of The 12th Language Resources and Evaluation Conference (LREC)},
    month = may,
    year = {2020},
    pages = {6532--6541},
    address = {Marseille, France},
    publisher = {European Language Resources Association (ELRA)},
    url = {https://www.aclweb.org/anthology/2020.lrec-1.804},
    ISBN = {979-10-95546-34-4},
  }

@inproceedings{si2022towards,
  title={Towards speaker age estimation with label distribution learning},
  author={Si, Shijing and Wang, Jianzong and Peng, Junqing and Xiao, Jing},
  booktitle={ICASSP 2022-2022 IEEE International Conference on Acoustics, Speech and Signal Processing (ICASSP)},
  pages={4618--4622},
  year={2022},
  organization={IEEE}
}

@INPROCEEDINGS{Dareeniassp25,
  author={Alharthi, Dareen and Zamani, Mahsa and Raj, Bhiksha and Singh, Rita},
  booktitle={ICASSP 2025 - 2025 IEEE International Conference on Acoustics, Speech and Signal Processing (ICASSP)}, 
  title={Tessellated Linear Model for Age Prediction from Voice}, 
  year={2025},
  volume={},
  number={},
  pages={1-4},
  doi={10.1109/ICASSP49660.2025.10888079}}

@article{reubold2010vocal,
  title={Vocal aging effects on F0 and the first formant: A longitudinal analysis in adult speakers},
  author={Reubold, Ulrich and Harrington, Jonathan and Kleber, Felicitas},
  journal={Speech communication},
  volume={52},
  number={7-8},
  pages={638--651},
  year={2010},
  publisher={Elsevier}
}

@article{vaswani2017attention,
  title={Attention is all you need},
  author={Vaswani, Ashish and Shazeer, Noam and Parmar, Niki and Uszkoreit, Jakob and Jones, Llion and Gomez, Aidan N and Kaiser, {\L}ukasz and Polosukhin, Illia},
  journal={Advances in neural information processing systems},
  volume={30},
  year={2017}
}

@book{schotz2006perception,
  title={Perception, analysis and synthesis of speaker age},
  author={Sch{\"o}tz, Susanne},
  volume={47},
  year={2006},
  publisher={Lund University}
}

@article{skoog2015can,
  title={Can you hear my age? Influences of speech rate and speech spontaneity on estimation of speaker age},
  author={Skoog Waller, Sara and Eriksson, M{\aa}rten and S{\"o}rqvist, Patrik},
  journal={Frontiers in psychology},
  volume={6},
  pages={978},
  year={2015},
  publisher={Frontiers Media SA}
}

@article{humayun2024review,
  title={A review of social background profiling of speakers from speech accents},
  author={Humayun, Mohammad Ali and Shuja, Junaid and Abas, Pg Emeroylariffion},
  journal={PeerJ Computer Science},
  volume={10},
  pages={e1984},
  year={2024},
  publisher={PeerJ Inc.}
}

@article{rubin1992nonlanguage,
  title={Nonlanguage factors affecting undergraduates' judgments of nonnative English-speaking teaching assistants},
  author={Rubin, Donald L},
  journal={Research in Higher Education},
  volume={33},
  pages={511--531},
  year={1992},
  publisher={Springer}
}

@inproceedings{gao2024documenting,
  title={Documenting ethical considerations in open source ai models},
  author={Gao, Haoyu and Zahedi, Mansooreh and Treude, Christoph and Rosenstock, Sarita and Cheong, Marc},
  booktitle={Proceedings of the 18th ACM/IEEE International Symposium on Empirical Software Engineering and Measurement},
  pages={177--188},
  year={2024}
}

@inproceedings{mireshghallah2021not,
  title={Not all features are equal: Discovering essential features for preserving prediction privacy},
  author={Mireshghallah, Fatemehsadat and Taram, Mohammadkazem and Jalali, Ali and Elthakeb, Ahmed Taha Taha and Tullsen, Dean and Esmaeilzadeh, Hadi},
  booktitle={Proceedings of the Web Conference 2021},
  pages={669--680},
  year={2021}
}

@article{tomashenko2024voiceprivacy,
  title={The voiceprivacy 2024 challenge evaluation plan},
  author={Tomashenko, Natalia and Miao, Xiaoxiao and Champion, Pierre and Meyer, Sarina and Wang, Xin and Vincent, Emmanuel and Panariello, Michele and Evans, Nicholas and Yamagishi, Junichi and Todisco, Massimiliano},
  journal={arXiv preprint arXiv:2404.02677},
  year={2024}
}

@inproceedings{feng2025developing,
  title={Developing a Top-tier Framework in Naturalistic Conditions Challenge for Categorized Emotion Prediction: From Speech Foundation Models and Learning Objective to Data Augmentation and Engineering Choices},
  author={Feng, Tiantian and Lertpetchpun, Thanathai and Byrd, Dani and Narayanan, Shrikanth},
  booktitle={Interspeech2025},
  year={2025}
}

@inproceedings{lertpetchpun2025developing,
  title={Developing a High-performance Framework for Speech Emotion Recognition in Naturalistic Conditions Challenge for Emotional Attribute Prediction},
  author={Lertpetchpun, Thanathai and Feng, Tiantian and Byrd, Dani and Narayanan, Shrikanth},
  booktitle={Interspeech2025},
  year={2025}
}

\newpage
\appendix

\section{Details on the Datasets}
\label{appendix-datasets}

\subsection{Datasets Descriptions}
\paragraph{TIMIT~\citep{garofolo1993darpa}} The TIMIT Acoustic-Phonetic Continuous Speech Corpus was collected by the Massachusetts Institute of Technology (MIT), SRI International (SRI), and Texas Instruments, Inc. (TI). The dataset includes audio recordings from 630 speakers representing eight major American English dialects. Moreover, the dataset contains speaker traits information, including speaker sex and speaker age. More than half of the speakers in this dataset are male.

\paragraph{VoxCeleb~\citep{nagrani2017voxceleb}} The VoxCeleb dataset is a large-scale audio-visual dataset designed for speaker recognition. The dataset contains more than 1,000 speech recordings from YouTube. The speakers have various nationalities, ages, and professions. The dataset includes metadata about speaker identity and speaker sex. Although age information is not provided in the original dataset, researchers in \citep{hechmi2021voxceleb} have enriched the age information by matching the speaker profiles with online sources. 

\paragraph{CommonVoice~\citep{ardila2020common}} The Common Voice dataset is an open-source collection of voice recordings to support the development of speech recognition technologies across many languages and demographics. Each data sample is a short audio clip of a person reading a provided sentence, along with self-reported metadata such as the speaker's age (every 10 years), gender, accent, and language. We follow data processing procedures in CommonAccent~\citep{zuluaga2023commonaccent} to prepare the data for training the accent classifier. Moreover, we use the same data to train the age and sex classifier.

\paragraph{EdAcc~\citep{sanabria2023edinburgh}} The Edinburgh International Accents of English Corpus (EdAcc) includes dyadic conversational recordings to study ASR performance with different language backgrounds. The datasets include speakers with different L1 languages, such as L1-Indian languages and L1-Spanish. We repurposed this dataset for our accent classification task.

\paragraph{British Isles Speaker~\citep{demirsahin2020open}} The British Isles Speaker dataset includes high-quality audio recordings of English-speaking utterances from subjects from different language backgrounds within the British Isles. Specifically, the dataset consists of recordings of 120 participants who are reported to be native speakers of Irish, Scottish, Welsh, and different regions in England.

\paragraph{L2-Arctic~\citep{zhao2018l2}} The L2-Arctic corpus is a collection of English speech recordings produced by non-native speakers. This dataset was designed to facilitate research like accented speech recognition. It includes recordings from speakers with diverse linguistic backgrounds, such as Korean, Mandarin, and Spanish. 

\paragraph{VoxPopuli~\citep{wang2021voxpopuli}} The VoxPopuli dataset is a large-scale multilingual speech dataset based on the European Parliament event recordings. In this experiment, we focus on the accented English speech subset that includes 15 different L1 language backgrounds, like French and German.

\paragraph{Fair-Speech~\citep{veliche2024towards}} The Fair-Speech dataset is a collection of speed data with diverse speaker meta information including age, sex, education levels, and accents. While there is no detailed information regarding L1 English speakers, we chose to only model accents with the L2 speakers who provided the exact language background information. 

\paragraph{ESLTTS~\citep{wang2024usat}} The ESLTTS is a dataset collected to advance research in text-to-speech (TTS) applications, with a particular focus on non-native English speakers.  The dataset contains speech samples from 134 non-native English speakers.

\paragraph{Hispanic-English~\citep{hispanic_eng}} The Hispanic-English Database (LDC2014S05) consists of 30 hours of English and Spanish conversations and read speech by native speakers of Spanish. Specifically, the English sentence prompts were designed based on the TIMIT dataset. We solely use this dataset to train the accent classifier.

\paragraph{Nigerian-English} The Crowdsourced Nigerian English Speech Dataset includes high-quality recordings of Nigerian English sentences, contributed by volunteers from Lagos, Nigeria, and London. The speech samples in this dataset are labeled under the "Other" category when training the accent classifier.

\paragraph{ParaSpeechCaps~\citep{diwan2025scaling}
} The ParaSpeechCaps dataset is a newly released, large-scale, human-annotated dataset that includes diverse speaker and speech information. It contains detailed annotations for accent, voice quality, and speech expressiveness from three datasets: VoxCeleb\citep{nagrani2017voxceleb}, EARS\citep{richter2024ears}, and Expresso~\citep{nguyen2023expresso}. Since only approximately 1.6\% of the holdout data is from EARS, we rely primarily on VoxCeleb and Expresso for training models related to voice quality and speech expressiveness.

\paragraph{MSP-Podcast~\citep{lotfian2017building}
} The MSP-Podcast dataset contains podcast data from the Internet, featuring spontaneous speech with natural human emotion expressions. In this experiment, we use the MSP-Podcast v1.12 dataset, which is also used for the IS2025-SER challenge. The dataset includes both arousal/valence and categorical emotion annotations.

\paragraph{SEP-28K~\citep{lea2021sep}
} The SEP-28k (Stuttering Events in Podcasts) dataset is one of the largest publicly available datasets for modeling disfluencies in speech. The dataset comprises 3-second audio clips from public podcast recordings. The disfluent speech includes blocks, prolongations, sound repetitions, word repetitions, and interjections. During inference, audio is processed using a sliding window approach, with a window length of 3 seconds and a step size of 1 second.

\paragraph{FluencyBank~\citep{ratner2018fluency}
} The FluencyBank dataset is an effort within the larger TalkBank project. Particularly, this dataset is designed to support the study of speech disfluency. It includes recordings of both spontaneous conversations and reading tasks. In our work, we use the disfluency annotations provided by \citep{lea2021sep}.

\subsubsection{Data Split}

\paragraph{Speaker Age and Speaker Sex} For age and sex prediction, we use the standard split provided in VoxCeleb Enriched~\citep{hechmi2021voxceleb} and TIMIT for training, validation, and testing. Moreover, we use the train, validation, and test split created from the CommonAccent~\citep{zuluaga2023commonaccent} pipeline to model the Common Voice dataset.

\paragraph{Speaker Accent} We use the standard data splits provided for the EdAcc, TIMIT, and Fair-Speech datasets. For the Common Voice dataset, we follow the data processing pipeline from the CommonAccent~\citep{zuluaga2023commonaccent} to create train, validation, and test sets. For the remaining datasets, we partition the data by selecting 60\% of the speakers for training, 20\% for validation, and 20\% for testing. We used a fixed seed when creating the data splits to ensure reproducibility.

\paragraph{Voice Quality} We use the standard data splits from the ParaSpeechCaps~\citep{diwan2025scaling} dataset.

\paragraph{Speech Expressiveness} We use the standard data splits from the ParaSpeechCaps~\citep{diwan2025scaling} dataset.

\paragraph{Speech Emotion (Categorical Emotion and Arousal/Valence)} We use the standard data splits from the MSP-Podcast v1.12 dataset. We report the development set results in Table~\ref{tab:speaker_attribute_result} and Table~\ref{tab:speech_attribute_result}, given the limited chance in challenge submissions. We report the test set performance in Table~\ref{tab:abalation_existing}. 

\paragraph{Speech Flow} We use the standard data splits from the SEP-28K dataset \citep{lea2021sep}. We also use the splits provided from the SEP-28K dataset \citep{lea2021sep} for FluencyBank \citep{ratner2018fluency}.

\subsection{Training Data Distribution}

\begin{figure} {
    \centering
    \begin{subfigure}[b]{0.75\textwidth}
        \includegraphics[width=\linewidth]{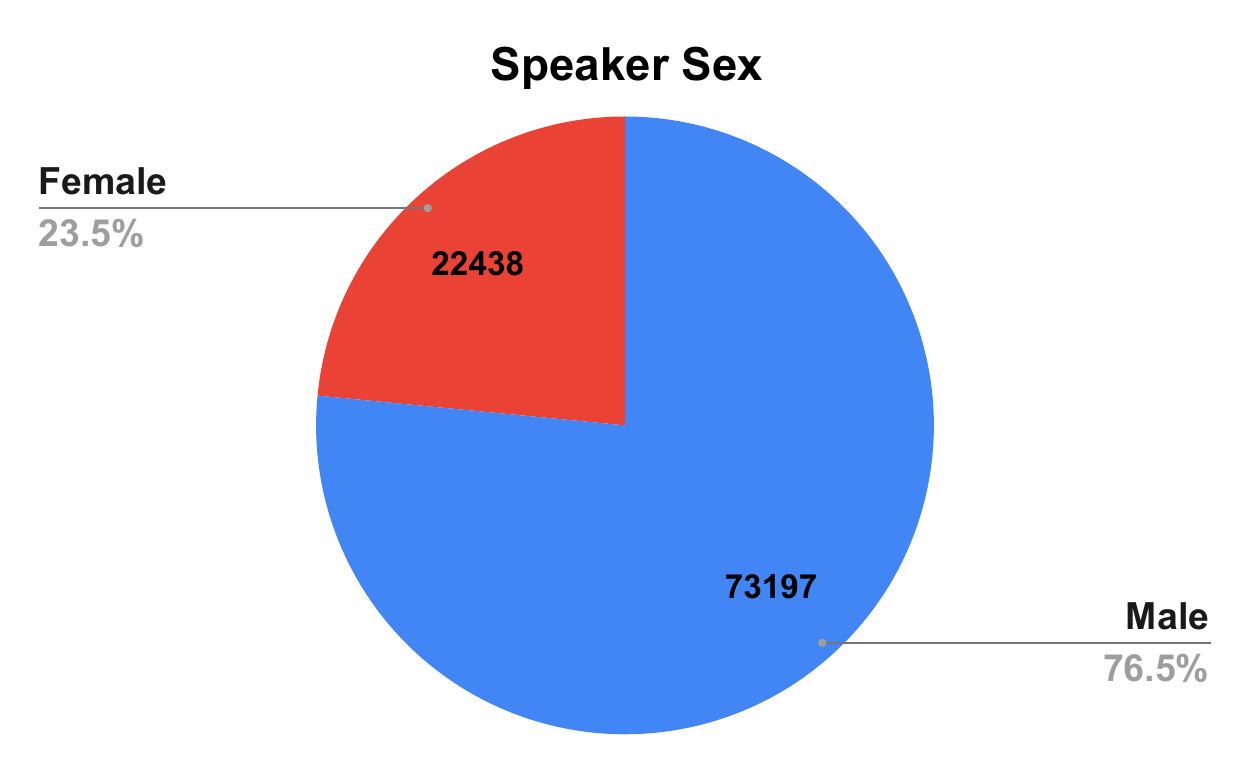}
        \caption{Speaker sex label distribution.}
    \label{fig:speaker_sex}
    \end{subfigure}

    \begin{subfigure}[b]{0.92\textwidth}
        \includegraphics[width=\linewidth]{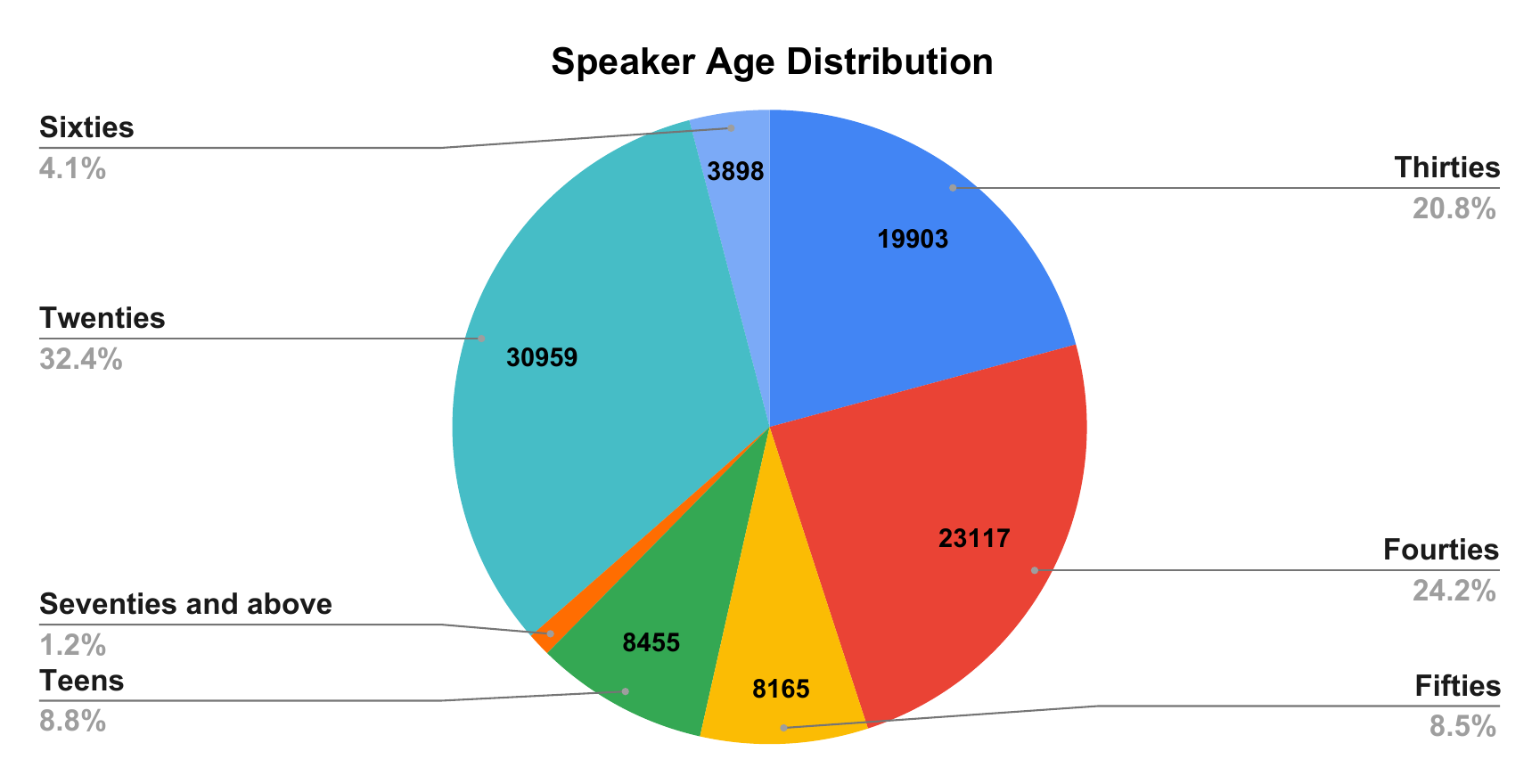}
        \caption{Speaker age label distribution in training data.}
    \label{fig:speaker_age}
    \end{subfigure}

} \end{figure}

\begin{figure} {
    \centering
    \includegraphics[width=\linewidth]{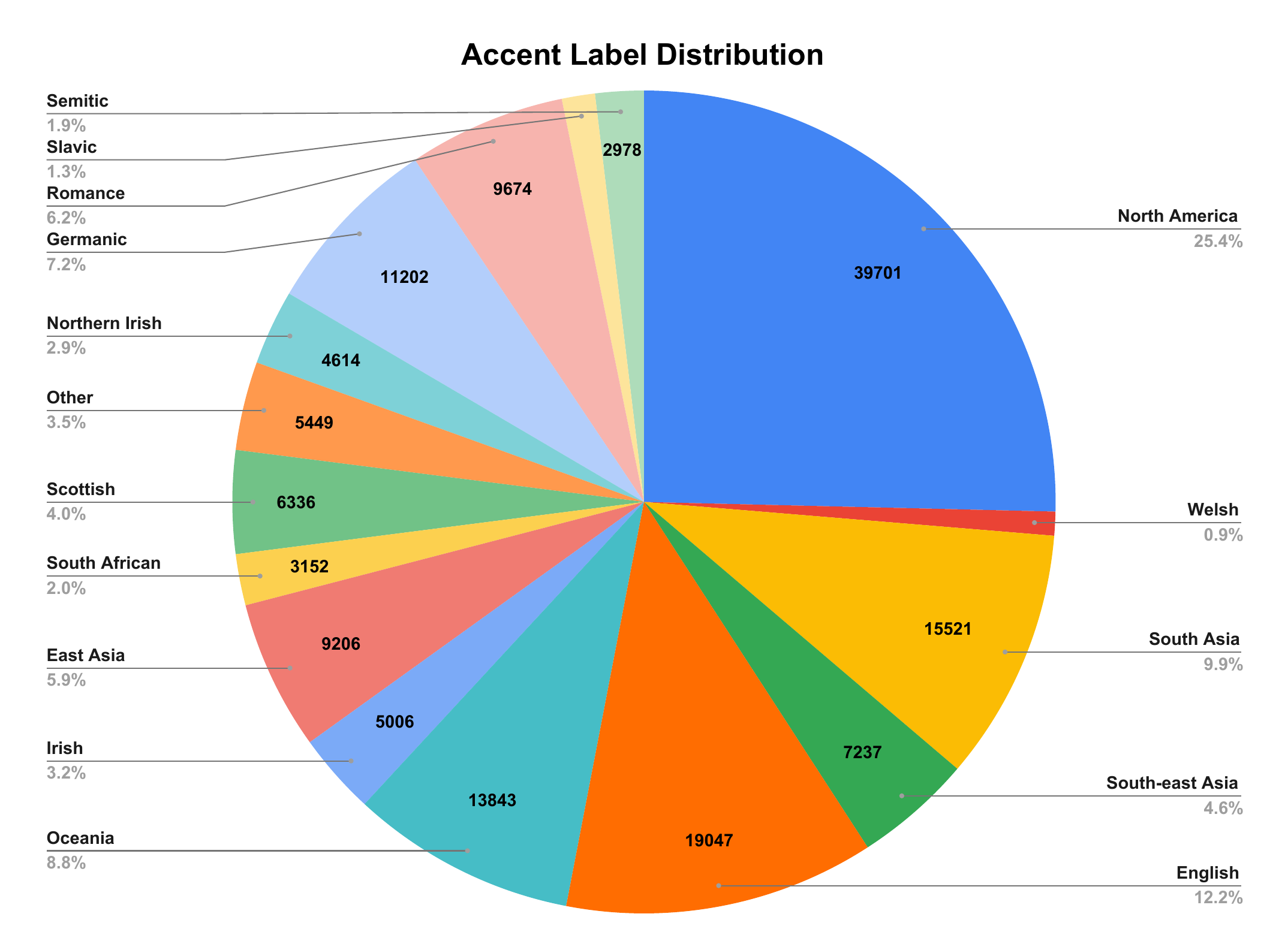}
    \caption{Accent label distribution in training data.}
    \label{fig:accent_label}
} \end{figure}

Given that we use the standard data splits from the ParaSpeechCaps dataset, the detailed label distribution for modeling voice quality and speech expressiveness can be found in \citep{diwan2025scaling}. Similarly, the detailed distribution of speech flow and emotion labels is reported in \citep{lea2021sep} and \citep{lotfian2017building}, respectively. Specifically, we plot the training label distribution of speaker sex, speaker age, and speaker accent in Figure~\ref{fig:speaker_sex}, Figure~\ref{fig:speaker_age}, and Figure~\ref{fig:accent_label}, respectively. Specifically, the sex distribution is male-dominant, given that a large portion of the data is from Common Voice. Regarding the speaker's age, more than half of the speech samples are from speakers between 30-60 years, and approximately 30\% of the data is from speakers below 30 years. Lastly, around 25\% of the data in \texttt{Vox-Profile} originates from North America, and about 12\% from speakers based in England.

\section{Details on the Models}
\label{appendix-models}

\noindent \textbf{ECAPA-TDNN}~\citep{desplanques2020ecapa} improves the traditional Time Delay Neural Network (TDNN) and x-vector architectures by integrating Squeeze-and-Excitation (SE) blocks, which improve channel-wise feature modeling. It achieves state-of-the-art performance across a wide range of speaker recognition and verification tasks. In this work, we utilize the ECAPA-TDNN model (\texttt{speechbrain/spkrec-ecapa-voxceleb}) available on Hugging Face.

\noindent \textbf{HuBERT}~\citep{hsu2021hubert} is a self-supervised learning model for speech representation learning. Building on top of Wav2Vec 2.0 \citep{baevski2020wav2vec}, HuBERT introduces the masked prediction of discrete units, leading to a robust representation of speech units. Specifically, HuBERT uses k-means clustering on MFCC or self-supervised features to create pseudo-labels as discrete units. In this paper, we use the HuBERT large (\texttt{facebook/hubert-large-ll60k}) from the Hugging Face.

\noindent \textbf{WavLM}~\citep{chen2022wavlm} expands on the Wav2vec 2.0~\citep{baevski2020wav2vec} pre-training objectives by incorporating masked speech and frame prediction. This model achieves competitive performance on popular speech-based downstream tasks, such as speaker recognition, speaker diarization, and speech recognition. In this paper, we use the WavLM large (\texttt{microsoft/wavlm-large}) from Hugging Face.

\noindent \textbf{Whisper}~\citep{radford2023robust} is a speech foundation model trained for ASR, language identification, and voice activity detection. The model was trained with more than half a million hours of audio data and achieved state-of-the-art ASR performance across multiple benchmark datasets. In this paper, we use the Whisper Tiny, Whisper Small, and Whisper Large-V3 from HuggingFace.

\section{Details on the Modeling Experiments}
\label{appendix-modeling}

All experiments were conducted using a fixed random seed to ensure reproducibility. During training, we applied several data augmentations to the input waveforms: Gaussian noise was added with a probability of 1.0, using an SNR range of 3–30 dB; time masking was applied with a probability of 1.0, using a masking ratio between 10\% and 15\%; time stretching was used with a probability of 1.0, with stretch rates ranging from 0.9 to 1.1; and polarity inversion was applied with a probability of 0.5.

In speaker sex, speaker age, voice quality, speech expressiveness, and speech flow classification experiments, we use a learning rate of [0.0001, 0.0005] and a training epoch of 10. Our experiments indicate that most models perform better with a learning rate of 0.0005. Moreover, we freeze the pre-trained model weights in all experiments. On the other hand, we use a learning rate of 0.0005 and a training epoch of 15 for training the speech emotion and speaker accent prediction systems. Specifically, for speech emotion recognition training, we also experiment with unfreezing the pre-trained weights in the WavLM Large model, following the IS25-SER challenge baseline. Empirically, we observe that unfreezing pre-trained weights for WavLM leads to improved performance in emotion prediction tasks. However, this is not consistently observed in training other classification systems. 

In all experiments, we use a batch size of 32 for training without LoRa and a batch size of 16 for training with LoRa. For all experiments, we report the utterance-level prediction performance except for the voice quality, given its high subjectivity. We apply the LoRa to the feedforward layer as suggested by our previous work~\citep{feng2023peft}.

When comparing with the existing literature, we use the HuggingFace model checkpoints of CommonAccent (Jzuluaga/accent-id-commonaccent\_ecapa) and the Robust Wav2Vec 2.0 age and sex prediction model (audeering/wav2vec2-large-robust-24-ft-age-gender). 

\subsection{Details on the Evaluation}
For age and sex classification, we compute both Macro-F1 and accuracy at the utterance level for each dataset, and report the unweighted average of these scores across three test datasets. Moreover, we report the utterance-level Macro-F1 and accuracy on the combined test set for accent classification, which includes speech samples from 11 sources as previously described. We avoid aggregating performance metrics across datasets due to substantial variability in label distributions. For example, the British Isles Speaker~\citep{demirsahin2020open} dataset only contains samples from the British Isles. For fluency and speech expressiveness prediction, we similarly report the unweighted average Macro-F1 scores between SEP-28K and FluencyBank.

For speech emotion recognition, we report utterance-level performance on the development set in Table~\ref{tab:speech_attribute_result}, as the \texttt{IS-25 SER Challenge} restricts evaluation frequency to be submitted on the test set. We present test set results based on limited evaluation submissions in Table~\ref{tab:abalation_existing}. For voice quality classification, we report speaker-level Macro-F1 scores. Specifically, for each speaker, we randomly sample five utterances and repeat this stratification process 20 times. The final score is computed as the average Macro-F1 across speakers. We then report the unweighted average of speaker-level Macro-F1 scores for VoxCeleb and Expresso. We exclude EARS from \texttt{ParaSpeechCaps} due to its limited number of samples in the holdout set.

\section{Details on the Resources}
\label{appendix-resouce}
All experiments are performed on one high-performance computing server with computing nodes on request. The computing node we request includes using NVIDIA A40 or NVIDIA V100 GPUs. All experiments require only a single GPU instead of multiple GPUs for training and inference. The longest training job is accent classification model training, taking approximately 60-80 GPU hours, given the variability of computing nodes allocated. The training time for the remaining classification models is all within 48 GPU hours. 

\subsection{Details on Speech Generation Task Evaluation}
\label{appendix-speechgentask}

For \texttt{FreeVC} and \texttt{VALLE-X}, we used the code and the checkpoint from here, respectively:\url{https://github.com/coqui-ai/TTS} and \url{https://github.com/Plachtaa/VALL-E-X}.
The selected test speakers and their self-reported dialectal regions are as follows: \texttt{p225 (England), p294 (USA), p326 (Australia), p234 (Scottish), p302 (Canada), p261 (Northern Ireland), p245 (Ireland), and p335 (New Zealand)}.
These speakers are the held-out test speakers of the pre-trained \texttt{FreeVC} model we used for inference. 3 utterances from each speaker were selected, forming each source-reference speaker pair with 9 (3 by 3) combinations. In total, 56 source-reference speaker pairs were chosen, adding up to 504 utterances. Cosine similarities reported in Table~\ref{tab:voice_cloning_similarity} are computed based on embeddings derived from the Vox-Profile accent prediction models.

\section{Details on Confidence Levels in Speaking Style Prompt Generation}
\label{appendix-confidence}

\begin{figure}[t] {
    \centering
    \begin{tikzpicture}

        \node[draw=none,fill=none] at (0,0){\includegraphics[width=0.5\linewidth]{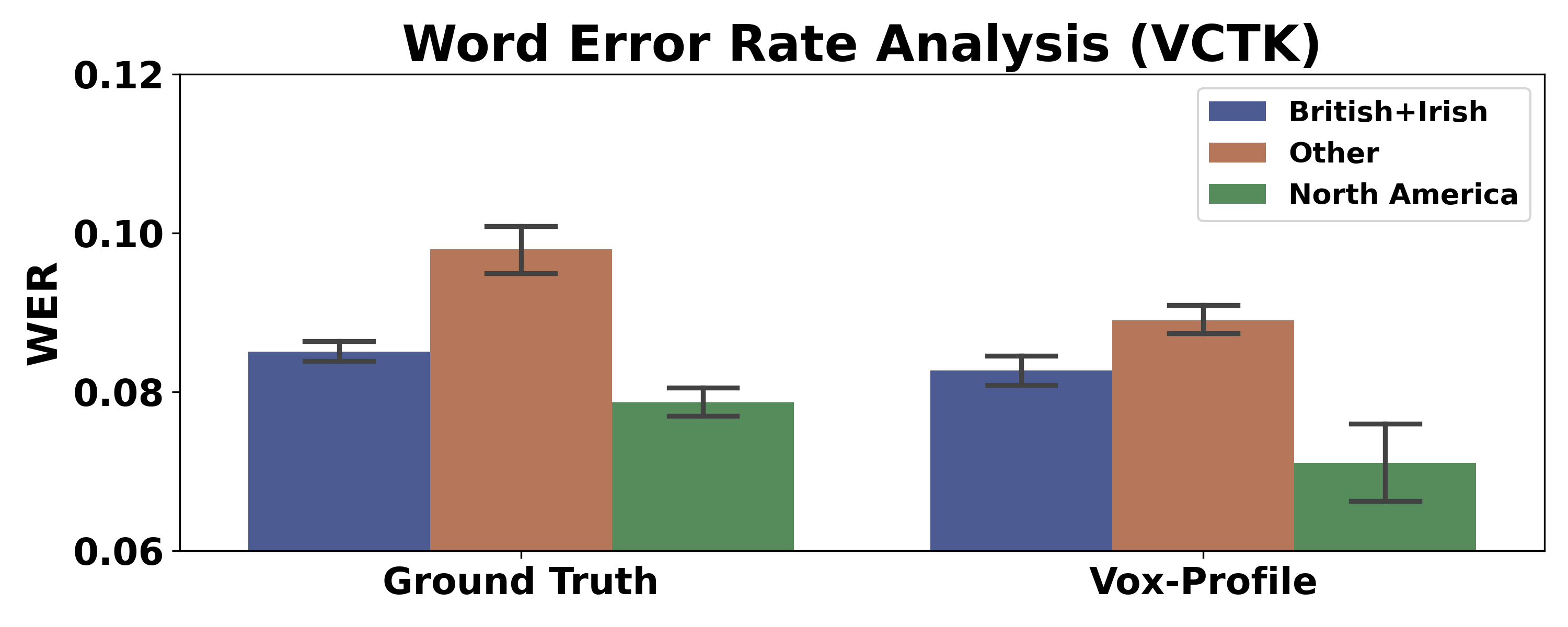}};

        \node[draw=none,fill=none] at (0.5\linewidth,0){\includegraphics[width=0.5\linewidth]{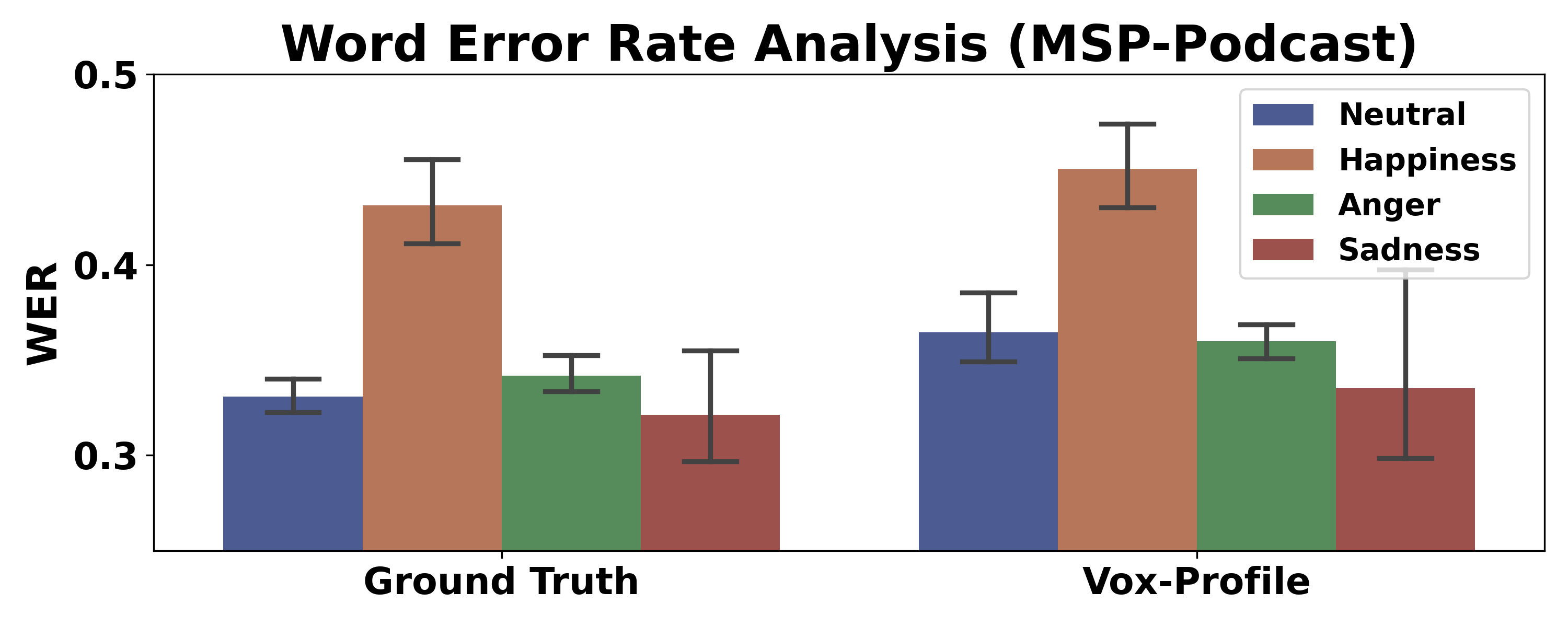}};
        
    \end{tikzpicture}
    \caption{ASR performance trends, grouped by ground truth and predicted labels by \texttt{Vox-Profile}. We measure WER, stratified by accent and emotion labels, with a threshold probability of 0.9 and 0.3, respectively. We observe similar performance trends between the predicted and ground truth trait labels as we observe in Figure~\ref{fig:vox_profile_asr}.}
    \label{fig:vox_profile_asr_higher_threshold}
} \end{figure}

\paragraph{ASR Experiments} For computing ASR performance using the speaker or speech labels generated by \texttt{Vox-Profile}, we use a threshold of 0.7 in filtering speech samples from the VCTK data. Moreover, we use a threshold of 0.25 to filter out emotional speech samples from the MSP-Podcast dataset. We would highlight that our results and conclusions are robust to threshold changes as presented in Figure~\ref{fig:vox_profile_asr_higher_threshold}. When we increase the threshold probability in filtering out the speech samples for comparing ASR performance, we observe the conclusion is similar as we observe in Figure~\ref{fig:vox_profile_asr} (With both cases of $p<0.05$). 

\begin{figure} {
    \centering
    \includegraphics[width=0.95\linewidth]{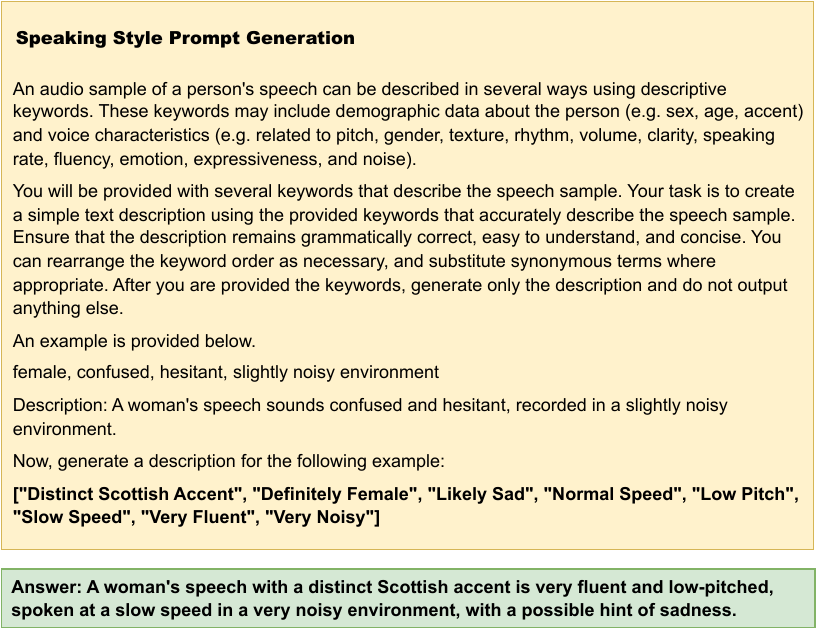}
    \caption{Prompt example for generating speaking style.}
    \label{fig:prompt_example}
} \end{figure}

\paragraph{Speaking Style Generation} A probability of 0.8 in sex classification is considered high confidence. For accent classification, we first apply the broad accent classifier given its strong performance in classifying speakers from North America, the British Isles, and Other regions or language backgrounds. A probability of 0.8 and 0.5 in broad accent classification is considered with high (distinct) and moderate (likely) confidence, respectively. When the accent is other, we further apply the narrow accent classifier to identify the precise accent labels. In narrow accent classification, a probability of 0.5 and 0.3 is considered with high (distinct) and moderate (likely) confidence, given the number of unique labels. If the maximum probability is below 0.3, we describe the accent as very hard to tell. In categorical emotion prediction, a probability of 0.4 and 0.3 is considered with high (very) and moderate (likely) confidence and otherwise was described as maybe. In arousal prediction, arousal scores below 0.2 and above 0.8 are described as calm and active. Similarly, valence scores below 0.2 and above 0.8 are described as negative and positive. For voice quality and speech expressiveness, only a probability above 0.8 is described, given the subjectivity in how human listeners perceive these traits. Similarly, a maximum probability above 0.8 in disfluency prediction will lead to the description of such a disfluency label. In addition to these predictive labels, we generate descriptions of pitch, speaking rate, and background noises. Specifically, an example of our prompt design and the corresponding response from GPT 4.1 is shown in Figure~\ref{fig:prompt_example}.

\begin{figure}[h]
  \centering
  \begin{subfigure}[b]{0.49\textwidth}
    \includegraphics[width=\textwidth]{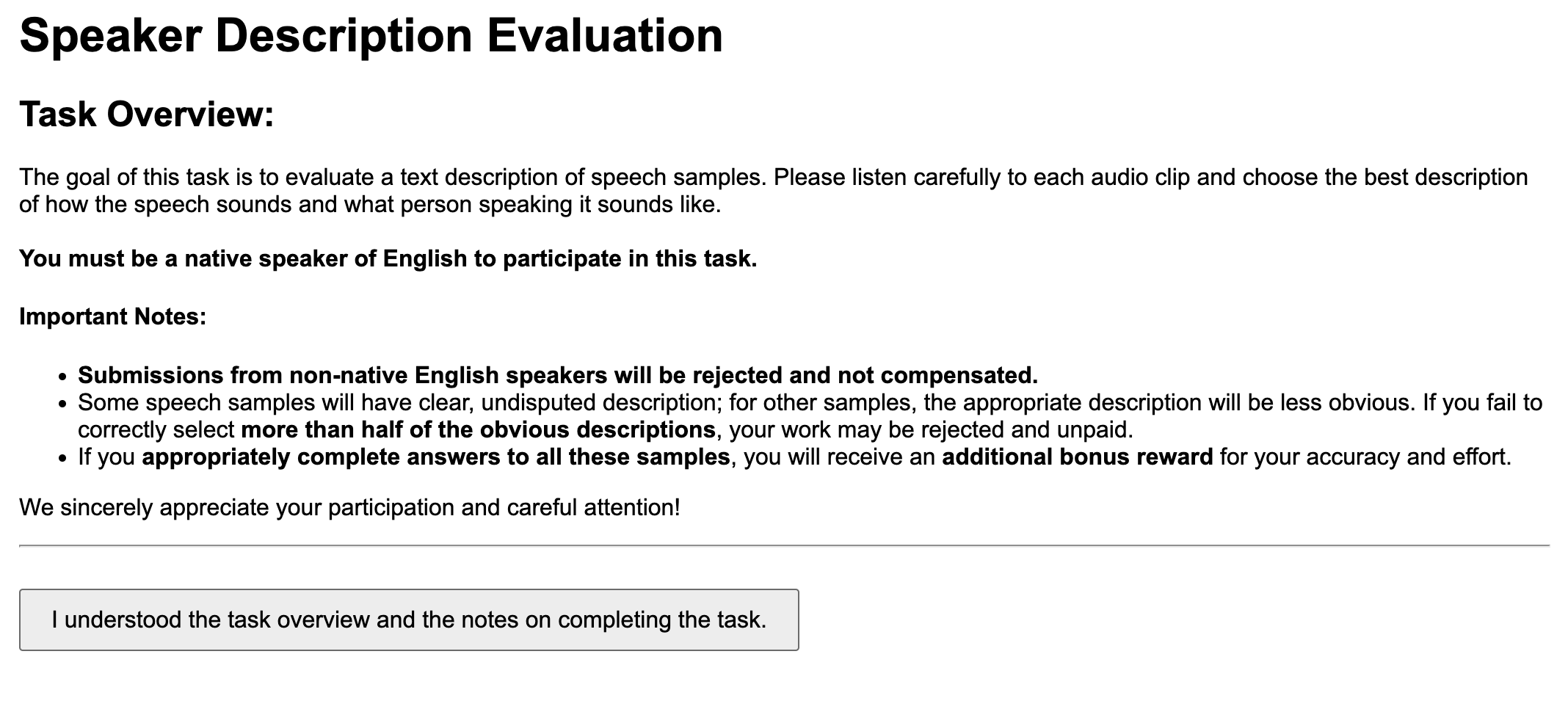}
    \caption{Task overview page}
    \label{fig:humaneval_instruction}
  \end{subfigure}
  \begin{subfigure}[b]{0.49\textwidth}
    \includegraphics[width=\textwidth]{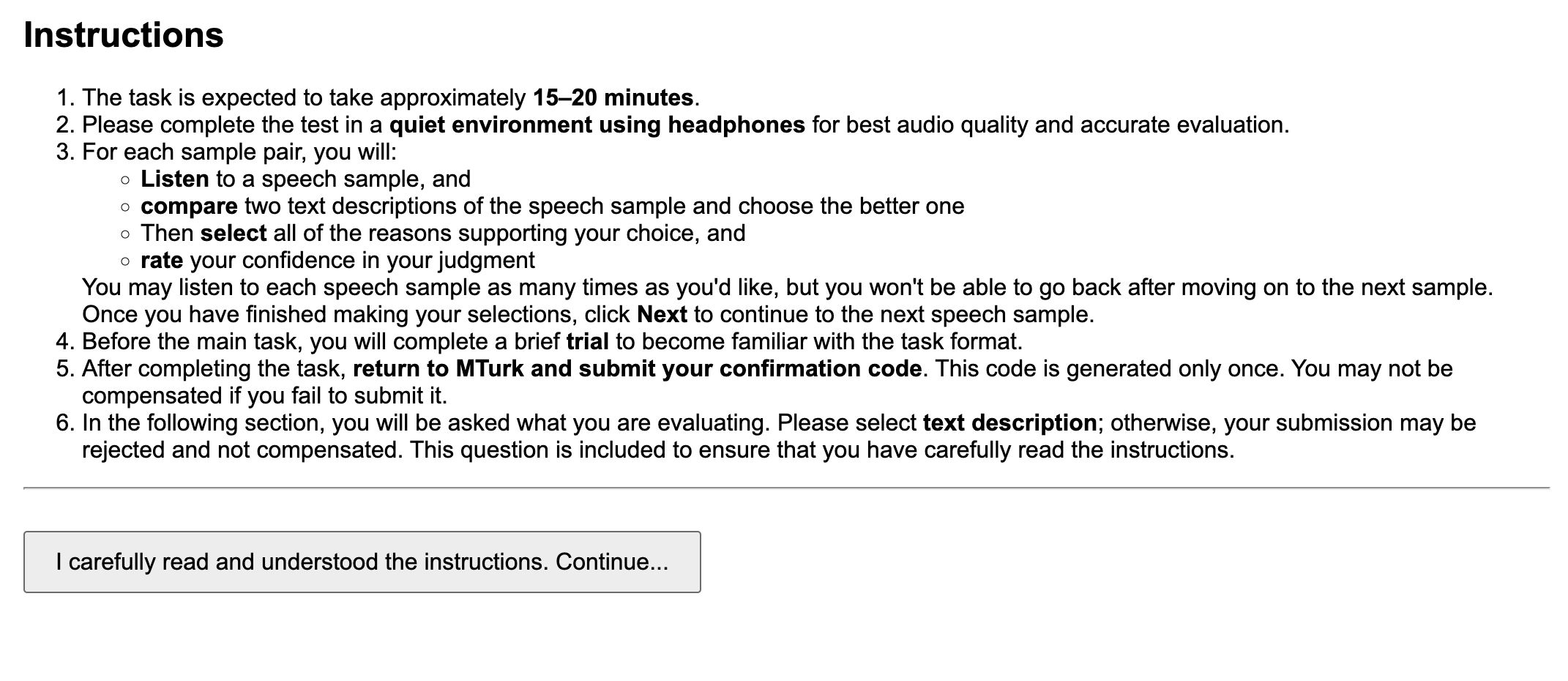}
    \caption{Instruction page}
    \label{fig:humaneval_sample}
  \end{subfigure}
    \begin{subfigure}[b]{0.99\textwidth}
    \includegraphics[width=\textwidth]{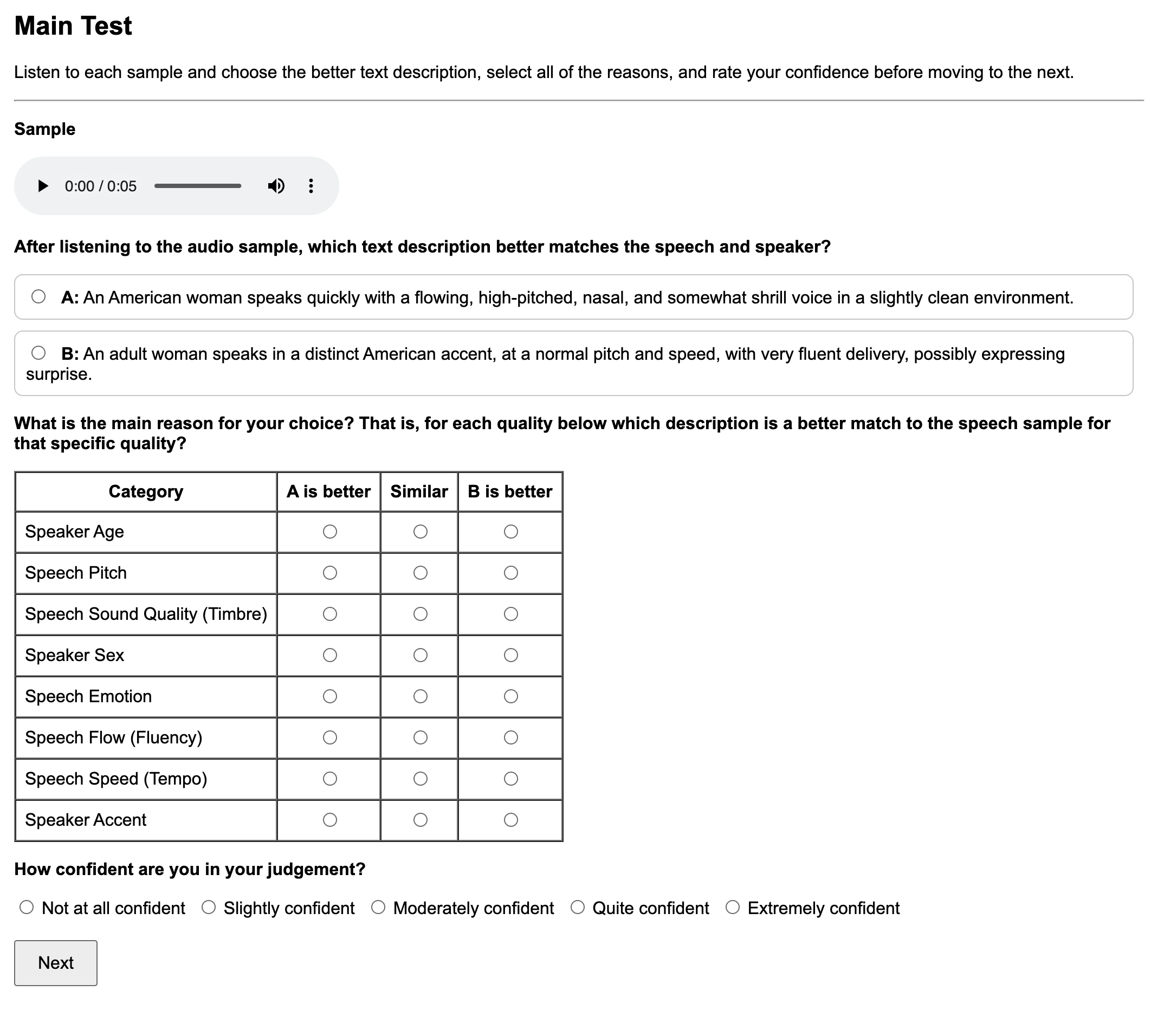}
    \caption{Main evaluation page}
    \label{fig:humaneval_response}
  \end{subfigure}
  \caption{Example screenshots of the human evaluation protocol.}
  \label{fig:humaneval}
\end{figure}

\section{Details on Text Description Human Evaluation}
\label{appendix-textdescripteval}

Human evaluators were first presented with a task overview and key instructions. Upon clicking the ``Next'' button, they received more detailed guidance on how to perform the evaluation, followed by a consent form. 
This guidance specified the trait dimension (e.g., voice quality, speech flow, accent) being evaluated and instructed raters to base judgments solely on the audio rather than the text.
To help them become familiar with the task, evaluators completed a brief trial session with a few sample items.
After the trial, participants proceeded to the main evaluation phase. Each evaluator was randomly assigned 10 samples from a total of 30, presented in random order. Each sample appeared on a separate page. Evaluators were allowed to listen to each sample as many times as desired, but once they advanced to the next page, they could not return to previous samples.
For each sample, participants were asked to choose the better of two text descriptions—one generated by our method and the other by \texttt{ParaSpeechCaps}~\citep{diwan2025scaling}. They then rated which description better captured specific traits, and these traits are presented in random order on every page. Finally, evaluators reported their confidence level for each judgment. Figure~\ref{fig:humaneval} presents example screenshots illustrating the human evaluation protocol. In total, we received 20 sets of evaluations with 200 utterance-level evaluations from approximately 10 pilot participants. Most of these participants are male and non-native English speakers.

\end{document}